# An SMDP-Based Approach to Thermal-Aware Task Scheduling in NoC-based MPSoC platforms

Farnaz Niknia, Kiamehr Rezaee, Vesal Hakami

**Abstract**— One efficient approach to control chip-wide thermal distribution in multi-core systems is the optimization of online assignments of tasks to processing cores. Online task assignment, however, faces several uncertainties in real-world systems and does not show a deterministic nature. In this paper, we consider the operation of a thermal-aware task scheduler, dispatching tasks from an arrival queue as well as setting the voltage and frequency of the processing cores to optimize the mean temperature margin of the entire chip (i.e., cores as well as the NoC routers). We model the decision process of the task scheduler as a semi-Markov decision problem (SMDP). Then, to solve the formulated SMDP, we propose two reinforcement learning algorithms that are capable of computing the optimal task assignment policy without requiring the statistical knowledge of the stochastic dynamics underlying the system states. The proposed algorithms also rely on function approximation techniques to handle the infinite length of the task queue as well as the continuous nature of temperature readings. Compared to related research, the simulation results show nearly 6 Kelvin reduction in system average peak temperature and 66 milliseconds decrease in mean task service time.

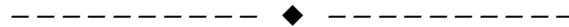

## 1 INTRODUCTION

### 1.1 Research Background

OLDER generations of computers (the so-called single-core systems) relied on a single processing unit, which could run only a single thread at any one time. Various parallelization techniques were applied on different levels, but despite the added complexity, increased power consumption, and heat generation, no significant improvement was achieved [1, 2]. The demand for higher performance and the shortcomings of single-core processors brought about the advent of Chip Multi-Processors (CMP). A CMP comprises several independent processing cores, as well as Network on Chip (NoC) routers and shared resources, managed by a single operating system (OS). Rather than relying on a large and powerful processing unit to improve performance, CMPs use several processing cores to run tasks in parallel to achieve more efficiency. Some types of CMPs further boost their energy-efficiency by also applying distinct voltage/frequency levels to processing cores depending on conditions and workloads [1].

Despite all the advantages associated with multi-cores, by integrating a large number of tiny transistors into a small chip, CMPs may have high power density and produce more heat, causing material fatigue, partial or complete failure, and shorter device lifetime. Although cooling equipment is used to prevent high temperature and thermal emergencies, specific workloads can aggressively use the processing resources, generating more heat than can be dissipated [2]. Moreover, not all processing units can be used simultaneously because of the high power and energy consumption of a CMP. This means that despite comprising several processing cores, the desired performance is not achievable. This phenomenon is called *dark silicon* [3]. One of the critical factors in increasing the power usage of a CMP is the temperature. High temperature increases the leakage power, thereby increasing the total energy consumption of the chip [2].

The dark silicone phenomenon on top of the damages looming from high temperature give rise to the dynamic thermal management (DTM) problem. The research on DTM aims at mitigating thermal emergencies and preventing thermal crisis in multi-core processors. Although failure-avoiding technologies like AMD's multi-point thermal control or Intel's Adaptive Thermal Monitor are being built into modern processors [2], the research is still ongoing, and there is a large body of work on thermal management in CMPs. In one broad taxonomy, one can classify the existing works into three groups: Dynamic Power Management (DPM), Dynamic Voltage and Frequency Scaling (DVFS), and Task assignment.

DPM [4-6] exploits the fact that the whole device or at least some of its parts may remain idle during specific time intervals and waste energy. These parts of the device (e.g. processing cores) can be turned off or switched to low power (sleep) states in idle periods in order to save

---

- *F. Niknia is with the Iran University of Science and Technology, Tehran, Iran. E-mail: niknia_farnaz@alumni.iust.ac.ir.*
- *K. Rezaee is with the Iran University of Science and Technology, Tehran, Iran E-mail: k_rezaee@comp.iust.ac.ir.*
- *V. Hakami is with the Iran University of Science and Technology, Tehran, Iran. E-mail: vhakami@iust.ac.ir.*

energy, while being brought back to active state when work becomes available. DPM can also be considered as a DTM technique given that it turns off the whole chip or hot cores in case of thermal emergencies in order to be cooled down [2]. Although DPM is advantageous in terms of energy and thermal management, frequent transitions between sleep and active states incur additional latency.

The idea behind DVFS [4, 7-9] is to reduce the power consumption of a processor via run-time adaptation of voltage and frequency (V-F) levels. DVFS relies on the fact that the dynamic power consumption of a system scales linearly with frequency and quadratically with voltage [10]. Despite its benefits, DVFS may result in increased execution time by applying lower frequency.

Both DPM and DVFS are considered "reactive" techniques as they typically take no measure to prevent thermal crises and are only activated when a predetermined temperature threshold is crossed [2]. Task assignment, on the other hand, can be considered as a "proactive" technique as it is aimed at preventing thermal emergencies rather than post-crisis activation. Also, unlike DPM and DVFS, it has no adverse effect on system performance. Depending on the position of the cores and their instantaneous temperature, tasks can be dispatched to processing cores located in different parts of the chip, thus distributing the heat in a chip-wide manner and reducing peak temperatures.

Our proposed scheme in this paper is a between DVFS and thermal-aware task assignment. We assume that random types of tasks may arrive at the system at stochastic time instants, and wait their turn in a queue to be assigned to processing cores. Tasks are assigned to cores by a dispatcher unit in the OS scheduler, while the voltage and frequency of the cores are regulated through a thermal manager component. Aside from handling stochastic task arrivals, our proposed thermal management scheme is a systematic methodology to face with other realistic (yet uncertain) factors. In particular, a recently assigned task may be randomly paired with an already running one, thus involving a number of NoC routers in addition to processing cores. As argued in [11] and [12], the uncertain NoC router involvements should also be accounted for by a task scheduler to more wisely control chip heat distribution.

In the sequel, to better highlight the research gaps and motivate our contributions, we review and categorize the most relevant literature.

### 1.2 Related Work

The simplest scheme that has addressed the thermal management problem in CMPs using task assignment is [13]. The authors have investigated the unwanted thermal cases, and presented a mechanism named Thermal-Aware Scheduling (TAS) that aims to moderate or even eliminate these thermal issues of CMPs. TAS has two forms: 1) the newly dispatched task is assigned to the coolest idle core. This is the simplest and easiest way to implement a thermal aware algorithm; 2) a cost is calculated for each idle core considering the temperature of individual cores as well as the temperature of cores in the neighborhood. Moving forward, several more advanced research works have investigated thermal-aware task assignment. Based on the availability/unavailability of the complete task graph at the time of assignments, task assignment approaches can be categorized into two groups: *batch* and *online*.

In *Batch* schemes, assignment decisions are made when all tasks have arrived to the system and are ready to run. *Online* methods, on the other hand, decide on the allocation of each incoming task individually while other tasks are currently running, entering the system or have not arrived yet. Therefore, unlike batch methods, at the time of each allocation, the system is in a different thermal condition which should be taken into account in assignments. Online methods can in turn be divided into two groups: *discrete-time* and *continuous-time*. With a discrete-time operation, each task is assigned at the beginning of fixed time intervals which leads to increased service time due to increased task waiting times.

Within another perspective, in many real-world systems, the scheduler faces several uncertainties including:

**Stochastic workload inter-arrival times:** Generally, workloads arrive at stochastic times, with typically no prior knowledge of their arrival instants [14]. It is noteworthy that methods related to the batch category do not consider arrival uncertainties. This is because of the fact that all tasks has arrived to the system before the assignment starts and no new tasks enter the system when deciding on the assignment of tasks.

**Stochastic workload characteristics:** in real-world problems, workload characteristics may not be known beforehand. This means that the execution times and the thermal impact induced by running each task are not given [14].

**Random pairing:** In general, the tasks arriving into the system can engage in IPC as clients, servers or both. A client is a process that requests a service from some other process. A server, on the other hand, responds to a client request. Many processes act as both a client and a server, depending on the situation. For example, a word processing task might act as a client in requesting a summary table from a spreadsheet process acting as a server. The spreadsheet process, in turn, might act as a client in requesting the latest inventory levels from an automated inventory control application [15]. In the absence of a deterministic and known task communication graph, IPC can be particularly challenging as each task (once assigned to a core), can randomly pair with some other currently running task. These random pairings can be due to several reasons; for example, IPC can be initiated by clipboard sharing when a user performs "cut, copy, and paste" operations, or it can arise from message passing for the exchange of randomly generated sensor data between a writer and a sensor program. IPC is assisted by and in fact relies on NoC routers, and since the pairings are random in general [16], [17], the routers to be involved in a communication are not known beforehand. Given that the NoC routers consume considerable power and produce a significant amount of heat compared to other chip components [18], it is vitally important to consider IPC-related uncertainties in our thermal-aware task assignments.

**Stochastic chip thermal profile:** the thermal impact of circuit components, the impact of thermal interface materials and cooling condition, make the thermal profile of the chip stochastic [18].

In the following, we review the related work in *batch* and *online* categories and with respect to the way the above-mentioned uncertainties have been dealt with:

- Within the category of *batch* methods, authors in [16] have observed that more heat will be dissipated if the processor reaches a high temperature earlier. Moreover, the authors have used a thermal model [17] in which the power consumption of a processor is calculated using the air ambient temperature and chip-dependent parameters. Exploiting this model, the authors indicate that in a particular time interval, a constant amount of energy is required to run a task. Therefore, to prevent temperature increase at the second half of the time interval, the temperature should rise as much as possible in the first half. Relying on these observations, they have proposed a greedy approach that runs the hottest job that does not violate the thermal threshold, at each step and in this way, they increase the operating temperature up to the threshold as quickly as possible. Also, to predict the future temperature of each core, a thermal predictor is presented using the hardware specifications, thermal sensor readings and the steady state temperature of an application which can be obtained by running each benchmark on each core until the temperature does not change anymore. Although this method is shown to outperform particular scheduling algorithms, it uses a deterministic thermal model to predict future temperature without considering thermal uncertainty. Moreover, selecting the hottest job, requires using the given thermal profile of tasks which violates the workload related (characteristics and pairing) uncertainties. [19] is a pioneer work that has addressed the tradeoff between energy consumption and thermal balance in a 2D mesh NoC architecture. There, a heuristic is presented based on multi-objective ant colony algorithm to explore task to core mapping space and find the pareto-optimal front that optimizes both energy consumption and hotspot temperature. There, the energy associated with the involved routers and links are taken into account to calculate the total energy consumption and to estimate the temperature of each tile. However, the authors in [12] have conveniently assumed that the task communication graph of the applications is given a priori, thus circumventing the uncertainties related to workload characteristics. The thermal model presented in [20] calculates the temperature difference between two tiles by only using the physical distance of tiles, chip-dependent parameters and the power consumption of the cores. This means that the thermal model is deterministic and thermal uncertainties are not considered. In [20], workload characteristics are assumed to be known a priori as well (i.e., the authors use an application task graph). A Kernighan-Lin bi-partitioning- based [21] approach is given in [22] to map the graph of an application on to a mesh-based NoC architecture with the aim of optimizing both communication cost and thermal-variance. Similar to previous approaches, this approach does not consider thermal and workload uncertainties. In [11], the authors have shown that the NoC routers have relatively small chip area and high power consumption compared to other on-chip components, thus they can potentially become hotspots. Accordingly, the authors analyze the importance of taking into account the NoC router power consumption in application mapping, and they considered the thermal effect of both cores and the NoC routers in their formulation. The authors have specifically addressed the tradeoff between temperature and network latency. Due to thermal correlation, reducing peak temperature requires placing high-power cores far from each-other, while the goal of reducing average latency may require these cores to be as close as possible. To address this issue, the authors have presented a temperature-aware partitioning and placement mapping approach using hierarchical bi-partitioning of the cores. However, the thermal model used to estimate the temperature of the cores is deterministic and does not take thermal uncertainties into account. Also, the algorithm exploits the task communication graph, sidestepping the issue of workload characteristics. Realizing that thermal correlation between heat sources may cause hotspots, [23] describes a thermal model in which the temperature of each location of a die depends on several factors such as on-chip heat-sources and their distances. Using this model, [23] searches the application graph and maps high communication flow tasks that do not communicate directly, on a column of the mesh topology and when there is no remaining tile, it turns to the right or left direction based on the minimal thermal correlation with the aim of reducing both peak temperature and communication cost. This method completely relies on a fixed a priori given application task graph. The work presented in [24] proposes a temperature-aware task scheduling approach for streaming applications on mesh-based NoC systems. There, a temperature model is built to estimate the temperature increase for processing a particular task. Using this model, the thermal profile of the tasks are first extracted at each V-F level. Then, for assigning each task, a priority value is calculated for each core using the estimated temperature increase of the task, the thermal effect of adjacent cores, the communication overhead and the location of the core on chip. Finally, the task is assigned to the core with the highest priority. However, the authors again assume that there is no uncertainty regarding the workload characteristics such as: the maximum number of clock cycles, the energy consumption, and temperature increase caused by each task execution. In both [25], [26], the authors rely on the fact that the cores located at the corners and the edges of the chip will cool faster if they become hot because they are surrounded with less cores. The maximum voltage and frequency level is used initially, then the V-F level is lowered at each step. When no lower V-F is available, tasks are migrated to cooler cores which causes displacement overhead. Authors in [27] exploit reinforcement learning to learn the best assignment policy for multi-threaded applications and adapt to workload changes. There, each action comprises assigning a task to a core and determining its working V-F level. The system states include thermal stress and aging that are calculated using performance counters and thermal sensor readings. To limit the state space, a discretization method is used to divide the working range of state elements into separate intervals and a representative is defined for each one. The values obtained from the thermal sensors and performance counters at each time, fall into particular intervals. Then, the original values are replaced with the relevant representatives. Although discretization limits the state-space and simplifies learning, it results in less learning precision. In addition, the size of intervals and their numbers affect the number of states and the learning algorithm, as well. Also, since standard Q-learning algorithm is used in [20] for learning the best decision making policy, storing Q-values for each state-action pair causes significant memory overhead. This is crucial, especially for embedded systems where the on-chip memory is limited [10]. In [28] an assignment approach is presented to reduce thermal hotspots and temperature gradients. The authors have defined several temperature thresholds for categorizing the processing cores based on their current temperature. Threads are also classified into three groups according to the temperature increase that they induce to a core. When a thread is ready to be assigned, according to its class, it is assigned to a core with proper temperature. Although, the classes of cores and threads are dynamically adjusted depending on new thermal conditions, it takes several iterations to determine a suitable group for each thread.

*Online* methods can also be further sub-classified as *discrete-time* and *continuous-time*:

- Among the *discrete-time* methods, in [25], the authors schedule the task queue under thermal and performance constraints such that a balanced temporal and spatial thermal profile is achieved. The scheduler obtains the temperature feedback of each individual core. Then, considering core temperature values and layout positions, a task queue length is assigned to each core and at next step, DVFS is applied to balance performance, temperature and task queue for each core individually. Applications are assigned to the cores based on the lengths of the queues, core layouts and the temperature readings from on-chip thermal sensors. This method takes arrival uncertainties into account by assuming that arrival intervals are Poisson distributed. Also, it does not exploit any workload characteristics for making assignment decisions. However, thermal uncertainties are not considered. Authors in [29] have implemented two simple heuristics: 1) selecting the coolest core for task allocation; 2) prioritizing cooler cores that have idle neighbors. They also presented an Adaptive Random scheme that assigns a probability to each core based on its thermal history. This value is increased if the core does not exceed a predetermined temperature, thus cores that keep temperature below the threshold for a longer time are more likely to be selected to run a task. It has been shown that Adaptive Random outperforms approaches that rely only on current temperature to make assignment decisions. However, it incurs memory cost for storing thermal history for each core [18]. Another shortcoming of this approach is that a list of jobs with their arrival times is provided for the scheduler beforehand which means that the stochastic task arrivals cannot be not accounted for. In [30] a scheduling algorithm is proposed for single-core processors to maximize the total number of completed tasks while keeping the temperature below a threshold. Time is divided into fixed intervals, and at each interval, the algorithm decides whether to schedule a task and which task to schedule. The future temperature of the system can be calculated assuming that the heat contribution of the tasks are given a priori. This algorithm has also been extended to multicore processors as well but the temperature of each core is calculated the same way as the single-core case, i.e., the heat transfer of adjacent cores are not considered. A similar strategy is used in [31] which has the same drawbacks as [30]. In [10], multi-threaded applications are executed for several iterations to learn the optimal assignment policy for each application with the aim of optimizing energy and temperature while addressing thermal aspects (peak temperature, average temperature, and thermal cycling). There, the thread allocation is separated from frequency scaling. The thread allocation is changed at long-term intervals using a greedy heuristic. A thermal overhead is then calculated after assigning each application using thermal cycling, average and peak temperatures. If the thermal overhead is reduced in comparison with the last allocation, the allocation is retained, otherwise, it is returned to the previous one. At the next step, the frequency selection is performed at every decision epoch using tabular Q-learning. Thermal and arrival uncertainties are ignored in this approach. In [26], for executing multi-threaded applications, a finite state machine is defined with five stages: *start*, *wait*, *read*, *calculate* and *assign*. In the *start* state, all variables are initialized, then the algorithm switches to the *wait* state where the system waits for a new time quota to run a new application. The *read* state is the next system state in which temperature values are collected from the on-chip thermal sensors and used in the *calculate* state where a cost matrix is built using system utilization data besides the location of each core. At last, the algorithm switches to the *assign* state and allocates tasks to cores according to minimal cost principle. However, arrival and thermal uncertainties are not considered. The problem of online task assignment is addressed more systematically in [18] where the assignment problem is formu-lated as a Markov Decision Process (MDP) [32]. The only system state is the vector of temperature values obtained from the on-chip thermal sensors, each action refers to assigning a task to a core, and the instantaneous reward is the temperature margin. Reinforcement learning is used to solve the MDP. The main drawback with [18] is that it is dubious with respect to the nature of time: On the one hand, the authors have mentioned that a task is allocated to an idle core as soon as a task arrival event occurs, which is reminiscent of a continuous-time operation; on the other hand, all their equations are based on a discrete-time MDP formulation. Also, since the task queue length is not explicitly considered as a state component in MDP, the uncertainties associated with task arrival and workload characteristics has not been captured in their formulation.

- As for continuous-time methods, in [33], the ready tasks are scheduled such that thermal emergencies are reduced in the presence of fixed ambient temperature. When a task is ready to run, a utilization factor is calculated, and then, core configurations are defined to select cores using a round robin fashion. When the temperature of a core exceeds the threshold, the configuration is changed. The well-known Global Earliest Deadline First (GEDF) algorithm is used as scheduling policy in [27]. This method, however, assumes periodic task arrivals, and does not consider arrival uncertainty. In [34], a method is presented to balance the usage of wireless links by avoiding congestion over wireless routers and to distribute temperature across the chip in many core system on chips. A 2D mesh wireless NoC is virtually divided into several regions such that each region contains a wireless router and each core falls into a region that has less hop count to its wireless router. Then, each application is mapped region by region in a round robin manner to balance thermal distribution while periodically rotating the Cartesian coordinate system to locally balance the thermal profile within each region. Finally, in each region, tasks of an application is mapped to cores with smallest indices. In [34], inter-application communications are given within a task communication graph while intra-application communications are considered to be highly dynamic. Also, there is no prior knowledge of applications' entrance time. Therefore, arrival and pairing uncertainties have been taken into account.

TABLE 1
RELATED WORK

| | Optimization Goal | Method | Category | | | Adaptive to | IPC | Uncertainty | | | |
|---|---|---|---|---|---|---|---|---|---|---|---|
| | | | Batch | Online DT | CT | | | Arrival | Workload characteristics | Pairings | Thermal |
| [16] | Throughput & utilization | Greedy | ✓ | | | × | × | × | × | × | × |
| [19] | Energy & temperature | Multi-objective ant colony | ✓ | | | × | ✓ | × | ×. | × | × |
| [22] | Communication & thermal-variance | Kernighan-Lin bi-partitioning | ✓ | | | × | ✓ | × | ×. | × | × |
| [11] | Temperature & latency | Hierarchical bi-partitioning | ✓ | | | × | ✓ | × | × | × | × |
| [23] | Temperature & communication | Mapping high communication flow tasks that do not communicate directly, on the non-adjacent tiles | ✓ | | | Thermal profile | ✓ | × | × | × | × |
| [24] | Peak temperature & temperature distribution | Prioritizing processing cores based on temperature | ✓ | | | Temperature | ✓ | × | × | × | × |
| [27] | The average temperature & thermal cycling | Tabular Q-learning | ✓ | | | Workload | × | × | ✓ | × | ✓ |
| [28] | Hotspots & temperature gradients | Categorizing processing cores and threads | ✓ | | | Temperature of cores & threads | × | × | ✓ | × | × |
| [25] | Balanced temporal and spatial thermal profile | Close loop control system | | ✓ | | Thermal profile | × | ✓ | ✓ | × | × |
| [29] | Temperature | Assigning probabilities to cores based on their temperature | | ✓ | | Thermal history of cores | × | × | ✓ | × | × |
| [30] | The total number of completed tasks | Predicting future temperature of the chip | | ✓ | | Workload & temperature | × | ✓ | ✓ | × | × |
| [31] | Weighted throughput | Using heat characteristics of the jobs | | ✓ | | Workload & temperature | × | ✓ | × | × | × |
| [10] | Energy & temperature | Heuristic and reinforcement learning algorithms | | ✓ | | Workload & temperature | × | × | ✓ | × | × |
| [33] | Thermal balancing | Selecting number of cores based on application utilization | | | ✓ | Workload | × | × | ✓ | × | ✓ |

| | Optimization Goal | Method | Category | | Adaptive to | IPC | Uncertainty | | | |
|---|---|---|---|---|---|---|---|---|---|---|
| | | | Batch | Online DT\|CT | | | Arrival | Workload characteristics | Pairings | Thermal |
| [26] | Heat distribution & ensuring the reliability | Finite state machine | | ✓ \| | Thermal profile | ✗ | ✗ | ✓ | ✗ | ✗ |
| [18] | Peak temperature | Q-learning | | ✓ \| | Thermal profile | ✓ | ✗ | ✗ | ✓ | ✓ |
| [34] | Global & local thermal distribution. | Dividing cores to regions - Round-robin scheduling | | \| ✓ | ✗ | ✓ | ✓ | ✗ | ✓ | ✓ |

## 1.3 Research gap and motivations

In many real-world systems, multi-core task processing is subject to various types of uncertainties. Reviewing the prior work, there is no scheme that accounts for all these uncertainties within a single unified framework. Probably the most neglected type of uncertainty is associated with IPC and task pairings; in fact, tasks of different applications may communicate with unknown patterns, and this type of inter-task interaction can be highly dynamic [34]. The random task pairings are particularly important in thermal-aware task assignment because they involve the NoC routers. These routers consume a significant amount of power in comparison with other on-chip components [4], and as they have a relatively small area, they can potentially become hotspots themselves [11]. Also, in most practical cases, there is no deterministic prior knowledge about the routers to be involved in each communication as the inter-application communication graph is not available beforehand [18, 34]. To the best of our knowledge, only [34] and [18] have partially taken task pairing uncertainties into account. Although they seem to be one step ahead of the other approaches, they also have several shortcomings. Aside from its dubious formulation (c.f., Section 1.2), the MDP model proposed in [8] accounts for IPC uncertainty only implicitly and through its indirect impact on the thermal profile of the chip. As with the work in [34], it is specialized for NoCs with wireless routers. Also, it fails to address other important types of uncertainties (e.g., workload characteristics), and provides no adaptability to thermal profile.

## 1.4 Overview of the Proposed Scheme and Contributions

Here, we give an overview of the proposed scheme together with a summary of our contributions in this paper:

- In order to account for the most common uncertainties prevalent in multi-core processing scenarios (including: stochastic task arrivals, unknown workload characteristics, random pairings, and unpredictable thermal profile of the chip), we systematically model CMP as a stochastic dynamic system. We also use the SMDP (Semi-Markov Decision Process) formalism [35] to formulate the online task assignment problem with the objective of maximizing the long-run average temperature margin of the chip. SMDP is among the fairly general variants of continuous-time optimization frameworks from the stochastic control theory. Since in a CMP, tasks arrive and depart in stochastic times, SMDP is a much more efficient choice compared to discrete-time formalisms (which would lead to an increase in task waiting times and degraded system performance). Each system state in our proposed SMDP is comprised of both discrete and continuous elements: the continuous element is the core temperature, and the discrete element includes the number of tasks inside the system (both in the task queue and in-service) as well as the idle/busy status of the processing cores. Each control action includes selecting a processing core to assign a task and determining its working V-F level. The operating system scheduler acts as the SMDP controller which is comprised of three units: *dispatcher* (assigns tasks to cores), *thermal manager* (applies working V-F level), and *thermal monitor* (reads the implemented on-chip thermal sensors).

In principle, the optimal scheduling policy can be calculated using model-based techniques for SMDPs such as dynamic programming algorithms [35]. However, these techniques rely on the availability of the prior knowledge of the system statistics (e.g., task arrivals, execution times, inter-task communication times, as well as the chip thermal dynamics). As these statistics cannot be realistically assumed to be available in all cases, to make up for this lack of knowledge, we propose a model-free scheme in which the scheduler agent is provided with the ability to sample-based experience and learning. Our proposed model-free solution is built on the well-known Q–learning algorithm from the MDP literature [35]. Standard tabular-based Q-learning, however, is suited particularly for discrete-state MDPs with relatively small state-action space dimension. Our assignment problem has a mixed continuous-discrete state space structure (corresponding to the temperature of processing cores and the number of tasks inside the system, respectively). Therefore, exploiting tabular Q-learning is not feasible for this problem as it would need infinite memory space. In addition, having a large state space needs more iterations for the convergence of Q-values. As such, we propose modifications of Q-learning which rely on function approximation techniques to handle the infinite length of the task queue as well as the continuous nature of temperature readings. In particular, we come up with two modified Q-learning algorithms as described below:

- **DVFS-Enabled:** In this variant, each action includes both assigning a task to a processing core as well as determining its working V-F level. To combat the curse of dimensionality associated with standard Q-learning, we exploit the notion of Radial Basis Functions (RBF) [36] to come up with a featureized representaion of states that include only the thermal features of the chip state. These features are obtained from on-chip thermal sensors that are implemented next to each core. RBFs are fed with the temperature values, and then, the summarzied thermal features are extracted from RBFs and used in function approximation for estimating the Q-values. The main advantage of the proposed DVFS-enabled scheme is reduced dynamic power dissipation, which is due to applying DVFS in addition to making intelligent task assignment decisions.

- **IR:** In this variant, each action is defined only as assigning a task to an idle core, without making any change to the working V-F level of the cores. Similary to our first scheme, in IR, function approximation with RBFs are used instead of tabular Q-learning, but the difference lies in the definition of the features. In addition to thermal features, in our IR scheme, we exploit several RBFs for extracting combined state-action features which include: 1) the Euclidean distance of a core from the chip center, 2) the Euclidean distance of a core from the hotspot, and 3) the ratio of the number of tasks whose communication paths are going to include the hotspot if they pair with the task assigned to some given core to the total number of tasks that may pair with the tasks running on the same given core. Our proposed IR scheme tries to assign tasks far from the chip center and the hotspot, effectively prefering the cores located at the chip corners and the edges. Compred to DVFS-enabled, IR needs a smaller number of learning parameters, making way for using a higher number of approximation functions. This can be exploited in larger mesh sizes where the number of learning parameters of DVFS-Enabled would grow unmanageably. As evidenced by simulations, IR outperforms DVFS-Enabled in 7x7 processors, indicating IR's superior scalablility.
- The processor and scheduler are simulated using multicore and thermal simulators including Sniper [37], Hotspot [38], McPAT [39], DSENT [40], Hotfloorpaln [38] which are used for simulating CMP, thermal profile, core power consumption, router power dissipation, and core floorplan, respectively. The simulation flow starts with simulating a CMP by sniper multicore simulator and running each Splash2 [41] benchmark on a single core to obtain its characteristics such as execution time. Then, using McPAT and DSENT, the power consumption of processing cores and NoC routers are obtained. Finally, Hotspot is fed with the floorplan of each processing core produced by hotfloorplan, and the power consumption of routers and cores. Hotspot thermally simulates the processor and produces the thermal profile of the whole CMP. At the final step, the simulation results are compared with two of the previous approaches. The results indicate that both the proposed DVFS-Enabled IR schemes reduce the long-run average peak temperature by 6 K and 5 K. Also, they reduce the average task service time by 110 and 40 milliseconds, respectively

## 2 SYSTEM MODEL

### 2.1 System Architecture

In this section, we first describe the system model for a multi-core processor including the processing cores, the operating system scheduler and the task queue. Then, we elaborate on the assumptions we make about the dynamics of the task arrivals, chip thermal profile, as well as the task pairings.

#### 2.1.1 Multi-core processor

We consider a multicore processor with an NoC-based mesh topology including $M$ processing cores and $M$ NoC routers (Figure 1). We denote by $\mathcal{M} = \{m_1, m_2, m_3, \ldots, m_m, \ldots, m_M\}$ the set of processing cores and by $\mathcal{R} = \{r_1, r_2, r_3, \ldots, r_m, \ldots, r_M\}$ the set of NoC routers. It is assumed that $r_m$ is the router connected to core $m_m$.

**Assumption 1:** *Each core is assumed to have a V-F regulator for adjusting its voltage and frequency individually. An example of a processor supporting core-level DVFS is the AMD Opteron "Barcelona" processor* [2].

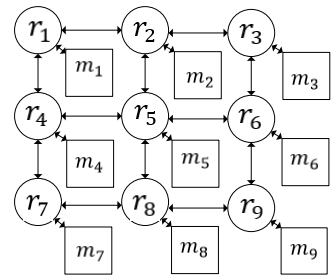

Fig. 1. A mesh-based multi-core processor [42]

#### 2.1.2 The operating system scheduler

In multi-core processors, the operating system scheduler is responsible for assigning tasks to the cores as well as for adjusting the working V-F levels. In order to manage the thermal profile of the multi-core processor, the scheduler needs to evenly distribute the heat across the chip-wide components by intelligently assigning tasks to cores located on different parts of the chip. As shown in Figure 2, the scheduler consists of three units [43]:

- Thread monitor: As electronic devices operate in specific normal temperature ranges [44], the current temperature of the cores must be taken into account when computing the task assignment policy in order to prevent thermal crises. As such, modern chips are equipped with thermal sensors to measure the temperature of the different parts of the chip. Here, we consider $M$ temperature sensors placed next to each processing core. The thermal monitor reads these thermal sensors, and sends the collected data to the thermal manager as well as to the dispatcher to be subsequently considered in task assignments. We denote the temperature readings at time $t$ by the vector $c(t) = (c_1(t), c_2(t), c_3(t), \ldots, c_m(t), \ldots, c_M(t))$ which includes M temperature values such that $c_m$ is the temperature of core $m_m$. Accordingly, one common thermal-sensitive performance measure is the so-called in-stantaneous temperature margin of the chip which we formally define as follows:

**Definition 1:** *The chip-wide temperature margin at time t is defined as:*

$$T_{mg}(t) \stackrel{\text{def}}{=} \max(T_{th} - T_p(t), 0) \quad (1)$$

where $T_{th}$ is the maximum tolerable temperature of the chip which is defined in datasheet of every electronic device and $T_p(t)$ is the peak temperature at time t which is calculated as:

$$T_p(t) = \max_m c_m(t) \quad (2)$$

- **Dispatcher:** This unit receives $c(t)$ from the thread monitor, and then (based on an assignment policy) selects an idle core for assigning a given task.
- **Thermal manager:** This unit receives $c(t)$ from the thread monitor, and then, it determines the operating voltage and frequency of the processing cores. We represent by $\mathcal{L} = \{l_1, l_2, l_3, \ldots, l_l, \ldots, l_L\}$ the set of applicable V-F levels to each processing core.

**Assumption 2:** *The task scheduler is implemented on a dedicated core.*

## 2.2 System uncertainties

A realistic multicore system is faced with many uncertainties (e.g., random task arrivals, uncertain IPC or thermal dynamics) which drastically influence the performance of any thermal-aware task assignment policy. In fact, in the presence of these uncertainties, the system undergoes a series of stochastic events over time, and the scheduler needs to adaptively adjust its decisions to cater not just for the current conditions, but also for the dynamic changes that are about to happen in the future. Hence, it is technically fair to say that the problem faced by a scheduler is sequential in nature, in the sense that every decision made in the present also has an impact on determining the long-run performance of the system. We elaborate further on this issue in Section 2.2 by giving a didactic example. Before that, we first discuss the most important types of uncertainties that need to be factored in our computation of a scheduling policy.

### 2.2.1 Stochastic task arrivals

In many real-world scenarios, the tasks entering a system are of different natures; for example, some tasks may heavily engage the arithmetic units (CPU-bound), while others mostly involve I/O operations (I/O-bound). In fact, executing tasks of various types results in different thermal profiles and takes different execution time. On top of that, different input sizes of an application also affects both its thermal footprint as well its execution time. Here, we assume that there are a total of $I$ different task types, each in turn associated with $N$ subtypes. In particular, let $\mathcal{I} = \{1,2,3,\ldots,i,\ldots,I\}$ denote the set of task types, and the set $\mathcal{N} = \{1,2,3,\ldots,n,\ldots,N\}$ be the set of all subtypes. Accordingly, we use the symbol $\tau_{i,n}$ to represent a task of type $i$ and subtype $n$. Also, the symbol $\zeta_{i,n,l}$ is used to indicate the pure execution time of $\tau_{i,n}$ by applying the operating level $L_l$ (i.e., without considering the inter-task communication time). It is further assumed that each task can only run on a single core. We denote by $\lambda_{i,n}$ the mean arrival rate of tasks of type $\tau_{i,n}$.

The key assumptions we make regarding the task model are as follows:

**Assumption 3:** *For $\forall i, n$, the task arrival process is Poisson with unknown parameter $\lambda_{i,n}$.*

**Remark 1:** *As all arrival processes is Poisson for $\forall i, n$, the aggregate arrival process into the multi-core system is also Poisson with unknown rate parameter $\lambda$.*

**Assumption 4:** *Each task entering the system has an exponentially distributed execution time (considering the communication time [18].*

**Assumption 5:** *There is no previous knowledge (neither acausal or statistical) about the task arrivals, the execution times nor the thermal impact of the tasks.*

### 2.2.2 Communication uncertainty

In order to capture the uncertainty associated with IPC and task pairings, we assume that each task entering the system has a chance of getting paired and communicating with another currently running task. This model of random task pairing can account for many real-world scenarios (e.g., clipboard sharing, random sensory data generation, etc.). Let $e_{i,n,j,y}$ denote the probability that task $\tau_{i,n}$ pair with $\tau_{j,y}$. The uncertain IPC model we envisage here is characterized by the following assumptions:

**Assumption 6:** *Each task can pair with just one of the running tasks at any one time.*

**Assumption 7:** *The duration of communication between any pair of tasks is exponentially distributed with unknown parameter $\xi_{i,n,j,y}$.*

**Remark 2:** *Given the deterministic nature of the pure execution times and the exponential distribution of the communication times, the total service time of the task $\tau_{i,n}$ paired with $\tau_{j,y}$ is an exponentially distributed random variable with the shifted parameter $\xi_{i,n,j,y} + \zeta_{i,n,l}$.*

**Assumption 8:** *The well-known xy routing algorithm [44] is used for determining the on-chip communication paths.*

**Assumption 9:** *The mean rate of data exchange is Poisson distributed with unknown parameter $\gamma_{i,n,j,y}$.*

### 2.2.3 Thermal uncertainty

The thermal impact of circuit components, the impact of thermal interface materials and cooling condition, make the thermal profile of the chip stochastic. We assume that all uncertainties of this kind can summarily be captured by an unknown perturbation parameter $\varrho$. Therefore, the next temperature values observed from thermal sensors, changes over time as follows:

$$c(t') = f(c(t), a(t), \varrho) \quad (3)$$

### 2.2.4 The arrival queue

The arriving tasks into the system are queued by the scheduler to be served in a continuous time FCFS fashion. Given the specifications of our system model, the arrival queue corresponds to an infinite-length multi-server queue with a Poisson arrival process, and exponentially distributed service time. We denote the occupancy state of the arrival queue at time $t$ by $q(t) \in \{1,2,3,...\}$.

**Assumption 10:** *The aggregate arrival rate $\lambda$ is assumed to be within the stability region of the queue. The queue stability region is a set $\Lambda$ entailing all arrival regimes (i.e., the $\lambda$ parameters), for which there is at least one scheduling policy $\pi$ under which the average length of the queue is bounded from above:*

$$\Lambda \stackrel{\text{def}}{=} \left\{ \lambda \in \mathbb{R}^+ \,\middle|\, \exists \pi: \lim_{T \to \infty} \frac{1}{T} \mathbb{E}^\pi \left[ \int_0^T q(t) dt \right] \leq \infty \right\} \tag{4}$$

Table 2 summarizes the notations used in our system model. Also, in Figure 2, we show a sample snapshot of the system at time $t$. The thermometers represent the relative temperature of the processing cores. These temperatures are first sensed by thermal sensors placed next to each core (shown with small circles next to the cores). The, the thread monitor collects these temperature readings and sends them to the dispatcher and to the thermal manager. According to a scheduling policy, the dispatcher assigns $\tau_{i,n}$ to some core $m_m$, and the thermal manager sets the working V-F level of core $m_m$ to some $l_l$. Of particular note is the continuous-time event-based operation of the scheduler. More specifically, the scheduling policy can be executed upon the arrival and departure events. In a fully-occupied system, as soon as the execution of a task is finished, and a core becomes idle, the head-of-line job (in case the task queue is non-empty) can be assigned to this newly idle core. Similarly, immediately upon arrival, a task can be assigned to an idle core (if there is any).

TABLE 2
SUMMARY OF NOTATIONS IN SYSTEM MODEL

| Description | Notation |
|---|---|
| Set of processing cores | $\mathcal{M}$ |
| Set of NoC routers | $\mathcal{R}$ |
| Number of processing cores, NoC routers and thermal sensors | $M$ |
| The $m$-th processing core | $m_m$ |
| The connected NoC router to core $m_m$ | $r_m$ |
| The chip thermal profile at time $t$ | $\mathbf{c}(t)$ |
| The temperature of core $m_m$ | $c_m(t)$ |
| Set of working voltage and frequency levels | $\mathcal{L}$ |
| The $l$-th voltage and frequency level | $l_l$ |

### 2.3 Problem statement

Before giving a formal definition of the thermal-aware task scheduling problem, in this section, we first give an illustrative example to further motivate our proposed solution in Section 3.

In Figure 3, we have illustrated a snapshot of the system where the tasks $\tau_{1,8}$, $\tau_{2,3}$, $\tau_{3,5}$, $\tau_{1,4}$, $\tau_{5,5}$, $\tau_{6,1}$ and $\tau_{1,7}$ are currently running on cores $m_1$, $m_7$, $m_6$, $m_2$, $m_4$, $m_3$ and $m_8$, respectively. At the same time, some new task $\tau_{2,8}$ enters the system, and the scheduler has the option of assigning $\tau_{2,8}$ to either of the idle cores $m_5$ and $m_9$. Assume that $m_5$ is currently cooler than $m_9$, and consider a myopic scheduler with an instantaneously greedy policy. Such a scheduler would assign $\tau_{2,8}$ to $m_5$. An unlucky run unfolds as follows: The newly assigned task $\tau_{2,8}$ gets paired with $\tau_{1,4}$ such that the NoC routers $r_2$ and $r_5$ would become involved in carrying out their IPC with a low in-jection rate. A few moments later, another task $\tau_{9,5}$ enters the system and choice-lessly gets assigned to the only idle core $m_9$. Now, it is quite likely that $\tau_{9,5}$ gets paired with $\tau_{1,8}$ such that the NoC routers $r_1$, $r_4$, $r_7$, $r_8$ and $r_9$ would become entangled with a high injection rate. In fact, as pairings could occur quite at random, the set of involved routers would not be predictable before task assignment. Given that the engaged NoC routers could generate significant heat, a foresighted scheduling policy that accounts for system uncertainties can bring about major improvements over a myopic policy that only considers the current temperature footprint to decide on its task-to-core mappings.

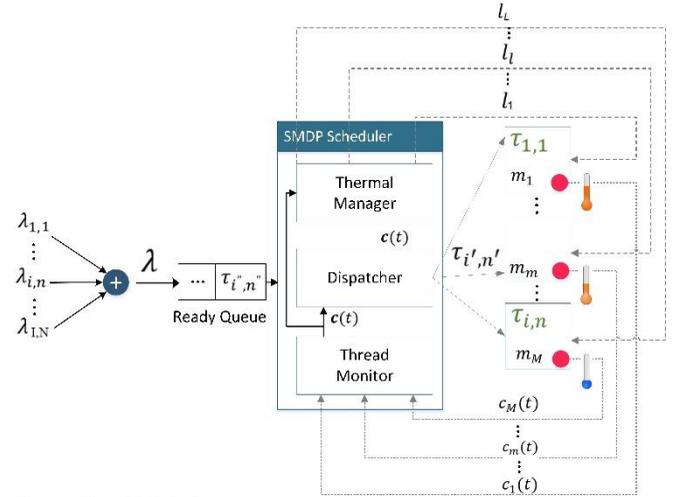

Fig. 2. The SMDP Scheduler

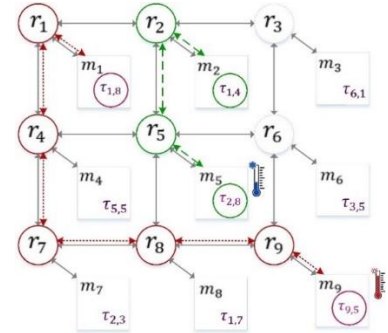

Fig. 3. An example task pairing scenario

In light of this simple example, one could perceive that the uncertainties associated with the task arrivals, task pairings, and the chip thermal dynamics can have a significant bearing on the thermal performance of a scheduling policy. In more technical terms, the multi-core processor gives rise to a stochastic dynamic system whose trajectory needs to be controlled with a state-dependent foresighted policy over time. Accordingly, unlike a myopic policy, it is not the instantaneous temperature margin of the chip that we seek to maximize. Rather, the scheduler needs to shape its course of actions so as to optimize a long-run thermal performance measure. Armed with this understanding, in Section **3**, we specify the structure of an adaptive scheduling policy and give a formal definition of our optimization objective that accounts for the uncertainties affecting the system.

## 3 PROBLEM FORMULATION

In the system model discussed in Section 2, the future thermal profile of the chip depends not only on the task assignment policy but also on important task characteristics such as task arrivals, pairings, and inter-task communication time. As these task characteristics are random in general, we need to cast the task-to-core assignment problem using a proper formalism from the stochastic optimization theory [45]. In particular, we use the framework of Markov Decision Processes (MDPs) [32] which is a standard formalism to express the decision problem faced by an agent operating in an uncertain environment. In an MDP, the decision made in each state affects the trajectory of the states to be visited by the system in the future, reflecting the sequential nature of the optimization problem. The objective is also expressed in a foresighted fashion which is to calculate an optimal policy that maximizes the long-run average reward accumulated by the agent. In general, the optimal policy in an MDP can be calculated in two different ways depending on the information availability assumptions [46]: *model-based* or *model-free*. Unlike model-based methods, a "model-free" method does not require prior statistical knowledge of the stochastic processes underlying the system, and this lack of knowledge is offset by equipping the device with the ability to learn and experience.

Also, most previous work on task scheduling (e.g., [10, 18, 27]) use discrete-time MDP in which tasks are assigned only at the start of fixed time quotas. This formulation, however, may lead to performance degradation in cases where several tasks are waiting inside the queue, and some processing cores are idle. The scheduler has to wait for a new time quota to assign the queued tasks. A further complication can arise with a discrete-time formulation, and it is the combinatorial nature of task-to-core assignments; in fact, in each decision epoch, the scheduler has no recourse but to consider all combinations of task-to-core assignments, which drastically adds to the complexity of the process.

Armed with this understanding, we propose a continues-time formulation based on the so-called semi-MDP or SMDP formalism [35] in which the scheduler is no longer limited to operate just at the start of fixed time quotas; instead, it has an event-based functionality; i.e., it is only invoked if an event (including a task arrival or departure) occurs. In case of a task arrival, if there is at least one idle core, one task will be assigned. Also, when a task departs, a core becomes idle. At this moment, if there is any task in the arrival queue, the scheduler will select an idle core to assign a task.

### 3.1 SMDP formalism

In this section, we formalize the task scheduling problem by casting it as an SMDP $\langle S, A, F, r, \pi, \bar{R}_\pi \rangle$, where $S, A, F, r, \pi$ and $\bar{R}_\pi$ denote respectively: the set of states, the set of control actions, system dynamics, instantaneous reward, the decision policy function, and the optimization objective. In the following, we elaborate on our formulation of the stochastic optimization problem.

#### 3.1.1 System states

The system state at time $t$ is denoted by $s(t) \in S$ which is comprised of three components:

$$s(t) = \{c(t), b(t), \kappa(t)\}c \qquad (5)$$

As before, $c(t)$ denotes the temperature readings at time $t$. Also, $b(t)$ is the state (idle/busy) of the processing cores:

$$b(t) \in \mathcal{B} = \{0,1\}^M \qquad (6)$$

In fact, $b(t)$ is a vector of 0s and 1s. For each core, the 0 value indicates that the core is idle, and 1 means that it is busy. $\kappa(t)$ represents the number of tasks in the system. In an infinite-capacity system, $\kappa(t)$ varies in the range of zero to infinity; i.e.,

$$\kappa(t) \in K = \{0,1,2,\dots\} \qquad (7)$$

#### 3.1.2 Actions

Let $\mathcal{A}(s)$ be the set of all feasible actions in state $s$. Further, $\mathcal{A} = \cup_{s \in S} \mathcal{A}(s)$. At each time $t$, a feasible action $a(t)$ at state $s(t) \in S$ is selected from the set $\mathcal{A}(s(t)) \subseteq \mathcal{A}$, and is comprised of two components:

$$a(t) = (m_m, l_l) \qquad (8)$$

In which $m_m$ is the selected core (possibly nil) to assign a task and $l_l$ is the chosen V-F level of that core. For example, in Figure 3, the system is considered to be at state $s(t_k)$ at time $t_k$. At this moment, a task enters the system and there may be several idle cores. The scheduler assigns the task at the head of the queue to the core $m_m$ and sets its working V-F level to $l_l$. Then, at $t_{k+1}$, some task departs and another core becomes idle. Now, if the ready queue is not empty, the scheduler assigns a waiting task to the core $m_{m'}$ with $l_{l'}$ V-F level. Finally, at $t_{k+2}$, a new task enters the system and is assigned to $m_{m''}$ with $l_{l''}$.

#### 3.1.3 System dynamics

Our formalization of the system's dynamics follows from the standard treatment of SMDPs given in [35]. Let $F_{ss'}(T, a)$ represent the conditional cumulative joint distribution of the transition time and the next system state conditioned on currently being in state $s$ and taking action $a$. In other words, $F_{ss'}(T, a)$ is the probability of transitioning from state $s$ to $s'$ by performing action $a$ with the transition time lasting less than or equal to $T$, i.e.,

$$F_{ss'}(T, a) = \mathbb{P}\left\{ \begin{array}{l} t_{k+1} - t_k \leq T, c_{k+1} \in \\ \hat{C}, b_{k+1} = \hat{b}, \kappa_{k+1} = \hat{\kappa} | \\ c_k = c, b_k = b, \kappa_k = \kappa, a_k = a \end{array} \right\} \qquad (9)$$

Moreover, the transition probability from state $s$ to $s'$ by taking action $a$ can be calculated as follows:

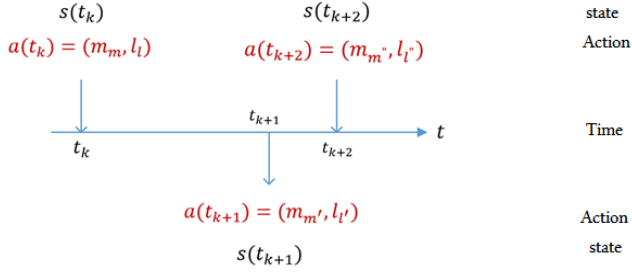

Fig. 3. Timeline

$$P_{ss'}(a) = \mathbb{P}(c_{k+1} \in \acute{C}, b_{k+1} = \acute{b}, \kappa_{k+1} = \acute{\kappa} | c_k = c, b_k = b, \kappa_k = \kappa, a_k = a) = \mathbb{P}(\acute{b}|b,a) \times \mathbb{P}(\acute{\kappa}|\kappa,a) \times \int_{\acute{C}} f_{cc'}(a) \, dc' \quad (10)$$

where $f_{cc'}(a)$ is the conditional transition probability density function of the chip temperature given the current temperature profile $c$ and action $a$. Note that all the three dynamics $\mathbb{P}(\acute{b}|b,a)$, $\mathbb{P}(\acute{\kappa}|\kappa,a)$, $f_{cc'}(a)$ as well as $F_{ss'}(T,a)$ are dependent on the statistics of the task arrival, task execution times and the inter-task communication times (c.f., Assumptions 3, 4 and 7). The thermal dynamics $f_{cc'}(a)$ also depends on the thermal perturbation function $f(c(t), a(t), \varrho)$ defined previously in Section 2.2.3.

Now, with the definitions of the $F_{ss'}(T,a)$ and $f_{cc'}(a)$ at hand, and given $a, s$ and $s'$, the conditional cumulative distribution of the transition time from $s$ to $s'$ by taking action $a$ can be computed as follows:

$$\mathbb{P}\begin{Bmatrix} t_{k+1} - t_k \leq T \ | c_k = c, b_k = b, \kappa_k = \kappa, c_{k+1} \\ \in \acute{C}, b_{k+1} = \acute{b}, \kappa_{k+1} = \acute{\kappa}, a_k = a \end{Bmatrix} = \frac{F_{ss'}(T,a)}{P_{ss'}(a)} \quad \mathbb{E}\{T|s,s',a\} = \int_0^\infty T \frac{dF_{ss'}(T,a)}{P_{ss'}(a)} \quad (11)$$

Also, given $a, s$ and $s'$, the conditional expected value of the transition time, i.e., $\mathbb{E}\{T|s,s',a\}$ can be calculated as:

$$\mathbb{E}\{T|s,s',a\} = \int_0^\infty T \frac{dF_{ss'}(T,a)}{P_{ss'}(a)} \quad (12)$$

Where $dF_{ss'}(T,a)$ is the derivative of the conditional cumulative joint distribution of the transition time and the next state.

Now, for each $s = (c, b, \kappa)$ and $a \in \mathcal{A}(s)$, the expected value of the transition time $\bar{T}_s(a)$ from state $s$ by taking action $a$ is calculated as:
$\bar{T}_s(a) =$

$$\sum_{\acute{b} \in \mathcal{B}} \sum_{\acute{\kappa} \in \mathcal{K}} \begin{bmatrix} \mathbb{P}(\acute{b}|b,a) \times \mathbb{P}(\acute{\kappa}|\kappa,a) \\ \times \int_{\acute{C}} f_{cc'}(a) \, \mathbb{E}\{T|s,s',a\} dc' \end{bmatrix} \Rightarrow \bar{T}_s(a) = \sum_{\acute{b} \in \mathcal{B}} \sum_{\acute{\kappa} \in \mathcal{K}} \int_{\acute{C}} \int_0^\infty T \, dF_{ss'}(T,a) dc'. \quad (13)$$

### 3.1.4 The instantaneous reward

In our thermal management problem, decisions should be made in a way as to reduce the average peak temperature of the system in the long-run. Instantly, the temperature of all processor components is measured. Then, the on-chip unit which has the maximum temperature is considered as the hottest spot and its temperature is marked as the system's peak temperature. Reducing the peak temperature of the system means maximizing the temperature margin. Therefore, the instantaneous reward $r(s(t), a(t))$ accrued in state $s(t)$ by performing action $a(t)$ at time $t$ is defined as:

$$r(s(t), a(t)) = T_{mg}(t). \quad (14)$$

### 3.1.5 Policy

In an SMDP-based formulation, we seek for the optimal policy as the solution of the optimization problem. In fact, each candidate policy function is a mapping which specifies what action should be taken by the decision maker in each state of the system. More formally, policy is defined as a mapping of the form:

$$\pi: \mathcal{S} \to \mathcal{A}, \quad (15)$$
$$\text{with } \pi(s) \in \mathcal{A}(s), \forall s \in \mathcal{S}.$$

### 3.1.6 The systematic goal (The average long-term reward)

In principle, the average of the instantaneous rewards obtained by following a given decision policy $\pi$ over an infinite-length time horizon represents the performance of the policy $\pi$ [35]. More formally, the long-term average "one-stage" reward for the continuous-time problem is calculated as follows [35]:

$$\rho^\pi = \lim_{T \to \infty} \frac{1}{T} \mathbb{E}_\pi \left\{ \int_0^T r(s(t), a(t)) dt \right\}, \quad (16)$$

where it is assumed that at each time instant $t$, the action $a(t)$ is drawn from the probability distribution $\pi(s(t))$. Now, the equation (16) is equal to: [35]

$$\rho^\pi = \lim_{N \to \infty} \frac{1}{\mathbb{E}_\pi\{t_N\}} \mathbb{E}_\pi \left\{ \int_0^{t_N} r(s(t), a(t)) dt \right\}, \quad (17)$$

where $t_N$ is the completion time of the $N$-th transition.

Therefore, our goal is to compute an optimal adaptive policy $\pi^*$ for assigning the tasks and selecting V-F levels for the cores such that:

$$\pi^* = \arg\max_\pi \rho^\pi \triangleq \max_\pi \lim_{N \to \infty} \frac{1}{\mathbb{E}_\pi\{t_N|s_0 = s\}} \times \mathbb{E}_\pi \left\{ \sum_{k=0}^{N-1} \int_{t_k}^{t_{k+1}} r(s_k, \pi(s_k)) dt | s_0 = s \right\} \quad (18)$$

Following the treatment in [35], under the assumption that the stochastic system state process $\{s(t)\}_{t \geq 0}$ is an ergodic Markov chain, the long-term average reward $\rho^\pi$ is well-defined and its value is independent of the initial state $s_0$. Next, in Section 4.1, we discuss how the optimal policy $\pi^*$ can be approximated in the absence of the statistical knowledge of the system's stochastic dynamics.

The symbols used for our SMDP formalism are gathered in Table 3.

TABLE 3
NOTATIONS USED FOR SMDP FORMALISM

| Notation | Description |
|---|---|
| $b(t)$ | The state (busy/idle) of all processing core at time $t$ |
| $\kappa(t)$ | The number of tasks inside the system at time $t$ |
| $\mathcal{B}$ | The (busy/idle) state space of all cores |
| $K$ | The state space of the number of tasks in the system |
| $a(t)$ | The selected action at time $t$ |
| $\mathcal{A}$ | The set of all actions |
| $\mathcal{S}$ | The set of all system states |
| $s(t_k)$ | The state of the system at $t_k$ (the $k$-th transition time) |
| $poissrand(\lambda)$ | The number of tasks that enter the system according to a Poisson process with parameter $\lambda$ |
| $O(t)$ | The number of terminated tasks within $(t, t')$ |
| $\varrho$ | The thermal dynamics perturbation parameter |
| $F_{ss'}(T, a)$ | Conditional cumulative joint distribution of the transition time $T$ and the next system state $s'$ given the current state-action pair $(s, a)$ |
| $\kappa_k$ | The number of tasks inside the system at $k_{th}$ transition |
| $b_k$ | The state (busy/idle) of all processing cores at $k_{th}$ transition |
| $c_k$ | The temperature of all processing cores at $k_{th}$ transition |
| $s$ | The current system state |
| $s'$ | The next system state |
| $f_{cc'}(a)$ | The conditional transition probability density function of the chip temperature given the current temperature profile $c$ and action $a$. |
| $P_{ss'}(a)$ | The transition probability from $s$ to $s'$ by performing $a$ |
| $r(t)$ | The instantaneous reward |
| $\pi$ | Decision policy |
| $\Delta(\mathcal{A})$ | The set of all probability distributions over the action space |
| $\bar{R}_\pi$ | The long-term average reward |
| $t_k$ | Time instant of the $k_{th}$ transition |
| $t_N$ | The completion time of the $N$-th transition |
| $R(s, a)$ | "one-stage" expected reward |
| $\bar{T}_s(a)$ | The expected value of the transition time from state $s$ by performing action $a$ |
| $\pi^*$ | The optimal policy |

## 4 THE PROPOSED REINFORCEMENT LEARNING ALGORITHM

The computation of the optimal policy $\pi^*$ can be done using standard dynamic programming techniques for SMDPs such as the specialized value iteration or policy iteration algorithms discussed in [32]. A major drawback with these techniques is that they require the system's stochastic dynamics $P_{ss'}(a)$ and $F_{ss'}(T, a)$ be known in advance. The characterization of these dynamics in our case would require the complete knowledge of the task arrivals, execution times, inter-task communication times, as well as the chip thermal dynamics. As these statistics cannot be realistically assumed to be available in all settings, to compensate for this lack of knowledge, we propose a model-free scheme in which the scheduler agent is equipped with the ability to experience and learn. In the sequel, we explain our proposed learning algorithm for task scheduling in a multi-core processing system. As with all forms of MDPs, the road to synthesize a learning algorithm starts with basic Bellman equations which we discuss next.

### 4.1 Bellman equations

Before stating the Bellman equations, we first establish one new notation. We define for each pair $(s, a)$ the so-called "one-stage" expected reward $R(s, a)$, which corresponds to the instantaneous reward from taking action $a$ in state $s$ normalized by the mean time spent in the transition from $s$ to some next state, and can be calculated as follows:

$$R(s, a) = r(s, a)\bar{T}_s(a) \qquad (19)$$

Now, according to the standard treatment of "average-reward" SMDPs, to solve for the optimal policy $\pi^*$, each state of the system $s \in \mathcal{S}$ is first given a "value" $V^\pi(s)$ under each candidate policy $\pi$. In particular, $V^\pi(s)$ denotes the expected total "per stage values" which would be obtained starting from that state and following policy $\pi$. These "per stage values" are defined as the difference between the "one-stage" expected reward $R(s, \pi(s))$ specific to the pair $(s, \pi(s))$ and $\rho^\pi \bar{T}_s(\pi(s))$ which denotes the long-term "one-stage" expected reward corresponding to policy $\pi$. Following the derivation in [35], this value function satisfies Bellman equations of the form given below:

$$V^\pi(s) = R(s, \pi(s)) - \rho^\pi \bar{T}_s(\pi(s)) + \bar{V}^\pi, \quad \forall s \in \mathcal{S} \qquad (20)$$

in which $\bar{V}^\pi$ is the expected value to be obtained by following policy $\pi$ onward from state $s$, and can be defined as follows:

$$\bar{V}^\pi = \sum_{\acute{b} \in \mathcal{B}} \sum_{\acute{\kappa} \in K} \left[ \begin{array}{c} \mathbb{P}(\acute{b}|b, a) \times \mathbb{P}(\acute{\kappa}|\kappa, a) \\ \times \int_c f_{cc'}(a) V^\pi(s') dc' \end{array} \right] \qquad (21)$$

where $s' = (\acute{b}, \acute{\kappa}, c')$ symbolizes the next system state. Since the optimal policy maximizes the value of all states, if the policy followed by the agent is the optimal policy $\pi^*$, the Bellman equation will change to equation bellow:

$$V^*(s) = \max_{a \in A(s)} [R(s, a) - \rho^* \bar{T}_s(a) + \bar{V}^*(s')], \forall s \in \mathcal{S} \qquad (22)$$

in which $V^*(s)$ is the maximum value of state $s$, $\rho^*$ is the optimal long-term average "one-stage" reward and $\bar{V}^*(s')$ is defined as follows:

$$\bar{V}^*(s') = \sum_{\acute{b} \in \mathcal{B}} \sum_{\acute{\kappa} \in K} \left[ \begin{array}{c} \mathbb{P}(\acute{b}|b, a) \times \mathbb{P}(\acute{\kappa}|\kappa, a) \\ \times \int_c f_{cc'}(a) V^*(s') dc' \end{array} \right] \qquad (23)$$

Similarly, for each action in each state, a value is assigned using the Bellman optimality equation [35].

$$Q(s, a) = R(s, a) - \rho^* \bar{T}_s(a) + \bar{V}^*(s') \qquad (24)$$

So we have:

$$V^*(s) = \max Q(s, a) \quad \forall s \qquad (25)$$

Finally, it follows that:

$$Q(s,a) = R(s,a) - \rho^* \bar{T}_s(a) +$$
$$\sum_{b' \in B} \sum_{\kappa \in K} \begin{bmatrix} \mathbb{P}(b'|b,a) \times \mathbb{P}(\kappa'|\kappa,a) \times \\ \int_c f_{cc'}(a) \max_{a' \in A(s')} Q(s',a') dc' \end{bmatrix}, \quad \forall (s,a) \tag{26}$$

Once the *Q*-function is computed for each $(s,a)$ pair, the optimal policy will be calculated as $\pi^* = arg \max_{a \in \mathcal{A}(s)} Q(s,a)$ [47].

However, for calculating $Q(s,a)$ based on equation (26), the transition probabilities $P_{ss'}(a)$ and $F_{ss'}(T,a)$ are required. Ironically, we also need to know the optimal long-term average "one-stage" reward $\rho^*$ itself!

In the absence of such knowledge, the well-known *Q*-learning algorithm [47] can be used. *Q*-learning is an iterative approximation procedure in which the learning agent actually experiences in the environment, and exploits samples of instantaneous rewards and observations of next state transitions to approximate $Q(s,a)$ values. Also, it has been shown that the estimates of the *Q* value for some fixed $(s^*, a^*)$ pair can replace $\rho^*$ in equation (24) [35]. In particular, the continuous-time version of the *Q*-learning algorithm is expressed as follows [47]:

$$Q_{k+1}(s,a) = (1 - \alpha_k)Q_k(s,a) + \alpha_k \Big[ r(s,a,s') - Q_k(s^*,a^*)t(s,a,s') + \max_b Q_k(s',b) \Big], \tag{27}$$

where $r(s,a,s')$ denotes the immediate reward earned in transition from $s$ to $s'$ by performing action $a$. $t(s,a,s')$ is the actual time spent in this transition, and $\alpha_k$ is the learning rate which is calculated as follows:

$$a_k = \frac{A}{B+k} \tag{28}$$

where $A$ and $B$ are some pre-defined constants.

In standard *Q*-learning, a *Q*-value is stored for each $(s,a)$-pair in a two-dimensional array called the *Q*-table. Then, at each stage $k$, the *Q*-value for the current $(s,a)$ is updated according to equation (27). In our thermal-aware task scheduling problem, each system state is comprised of three components, viz. the state of processing cores, temperature values across the chip, and the number of tasks inside the system. In large-scale many-core systems, the state component associated with the idle/busy status of the cores would have a high dimension. Also, with a task queue of infinite capacity, the number of tasks inside the system can theoretically grow without limit. Most problematic though is the thermal profile of the chip which is a vector of sensor readings all of continuous real-valued nature. Therefore, the state space is infinitely large, and due to memory limit, using the standard form of *Q*-function with a *Q*-table is not practical. In general, when facing with continuous state values, two different methods can be used: *discretization* and *function approximation*. With discretization, the continuous elements of the state vector are discretized into several intervals. Then, a representative is defined for each one and is used instead of all values in that interval [7, 10, 27]. However, learning becomes much slower using discretization due to the need for more trials and errors [۴۸]. Moreover, the size of the intervals would drastically affect the learning precision. In the RL literature, the preferred way to combat the curse of dimensionality is to come up with a function approximation architecture. In the next section, we propose a novel approximation architecture suitable for our problem at hand.

### 4.2 Scaling the Q-learning Algorithm: The Proposed Function Approximation Architecture

To scale the *Q*-learning scheme, in this section, we propose a function approximation architecture for approximating state values using state similarities [36]. Intuitively, the more similar a pair of states are, the more similar their values should be [49]. Hence, the learner does not need to experience all system states to estimate their values and it can generalize its knowledge to unseen states with similar features, which can greatly increase the convergence rate. A common way to approximate the *Q*-table is to use a linear approximation of the form below [50]:

$$\hat{Q}(s,a) = \boldsymbol{\theta}^T \times \boldsymbol{\phi}(s,a), \tag{29}$$

where $\boldsymbol{\phi}(s,a)$ is a vector-valued feature function that maps the state-action values to a summarized feature space and $\boldsymbol{\theta}$ is a weight vector denoting the relative importance of each feature. We leave the details of the proposed $\boldsymbol{\phi}(s,a)$ to Sections 4-2-1 and 4-2-2, in which two versions of the *Q*-learning algorithms are introduced for the task scheduling problem. Hence, for now, we suppose that $\boldsymbol{\phi}(s,a)$ values are defined. Then, the *Q*-learning algorithm in Section 4.1 should be modified as follows: After each decision making in stage $k$, the vector $\boldsymbol{\theta}_k$ should be updated so as to minimize the mean square error ($E_k$) of the new perception ($Q_{k+1}$) and its approximated value ($\hat{Q}$) [36]:

$$E_k \triangleq [Q_{k+1}(s_k, a_k) - \hat{Q}(s_k, a_k)]^2$$
$$= [Q_{k+1}(s_k, a_k) - \boldsymbol{\theta}^T \times \boldsymbol{\phi}(s_k, a_k)]^2 \tag{30}$$

in which $Q_{k+1}(s_k, a_k)$ is calculated based on the most recent observation as follows:

$$Q_{k+1}(s_k, a_k) = r(s_k, a_k, s_{k+1}) - \hat{Q}(s^*, a^*) t(s_k, a_k, s_{k+1}) + \max_{b \in A(s_{k+1})} \hat{Q}(s_{k+1}, b) \tag{31}$$

For minimizing the error $E$, we use the gradient descent technique (with $\beta_k$ being the step size):

$$\nabla_{\boldsymbol{\theta}} E = -2\boldsymbol{\phi}(s_k, a_k)[Q_{k+1}(s_k, a_k) - \boldsymbol{\theta}^T \times \boldsymbol{\phi}(s_k, a_k)] \tag{32}$$

Then $\boldsymbol{\theta}_k$ is updated as follows:
$$\boldsymbol{\theta}_{k+1} = \boldsymbol{\theta}_k - \beta_k \nabla_{\boldsymbol{\theta}} E \tag{33}$$

By defining $\alpha_k \stackrel{\text{def}}{=} 2\beta_k$, we have:
$$\boldsymbol{\theta}_{k+1} = \boldsymbol{\theta}_k + \alpha_k (Q_{k+1}(s_k, a_k) - \hat{Q}(s_k, a_k))\boldsymbol{\phi}(s_k, a_k) \tag{34}$$

#### 4.2.1 Proposed (DVFS-Enabled)

As discussed in the previous section, each element of the vector $\boldsymbol{\phi}(s,a)$ is called a *feature*; $\phi_i(s,a)$ denotes the value of feature $i$ for state-action pair $(s,a)$. The feature function $\boldsymbol{\phi}: \mathcal{S} \times \mathcal{A} \to \mathbb{R}^g$ maps each pair $(s,a)$ to a vector of feature values. Finding the right feature function plays a key role in the success of our RL-based algorithm.

Our first suggestion is to first use standard Radial Basis Function (RBF) as the state-only feature function [36], and then construct the whole feature function $\boldsymbol{\phi}(s,a)$; in particular, the $i$-th component of $\boldsymbol{\phi}(s)$ is defined as:

$$\phi_i(s) = \frac{1}{\sqrt{2\pi\sigma^2}} e^{-||c-\omega||^2/2\sigma^2}, \tag{35}$$

where $\sigma$ is a constant reflecting the width of the features and $\boldsymbol{\omega}$ is an $M$-element vector that specifies centers of the features. In our context, each center $\omega_m$ denotes a value that lies within the normal temperature range of a processor (e.g., $\omega_m \in [330, 360]$ Kelvin). Also, $\boldsymbol{c} - \boldsymbol{\omega}$ calculates the difference between the thermal sensor readings and the normal temperature value. Typically, rather than letting $\omega_m$ choose values from a continuous range, each element $\omega_m$ is assumed to be chosen from one of $x$ discrete values. For example, if $x = 2$, then each feature will have 2 centers (e.g., $\omega_m \in \{340, 350\}$). Thus, we will have $g = x^M$ combinations for the vector $\boldsymbol{\omega}$, and a total of $g$ feature functions $\phi_i(s), i = 1, \ldots, g$ to estimate the value of each state. Now, with this scheme, the features are extracted only from the thermal component of the system state $s$. Also, the whole point in approximating the $Q$-table is to be able to correctly rank the actions in a given state. Therefore, if $a \ne a'$, features used for approximating $Q(s,a)$ should be different from features that are used for approximating $Q(s,a')$. Following the discussion in [50], given a pair $(s,a)$, this constraint is met by mapping $s$ to a vector of feature values $\phi_i(s)$, and then using these values in the corresponding slot for action $a$ while setting the feature values for the rest of the actions to zero.

The following example shows this mechanism for a system with 2 actions and 3 features per action. Hence 3x2 = 6 features are used for linear function approximation.

$$\phi(s) = \begin{bmatrix} \phi_1(s) \\ \phi_2(s) \\ \phi_3(s) \end{bmatrix} \Rightarrow \phi(s, a_1) = \begin{bmatrix} \phi_1(s) \\ \phi_2(s) \\ \phi_3(s) \\ 0 \\ 0 \\ 0 \end{bmatrix}, \phi(s, a_2) = \begin{bmatrix} 0 \\ 0 \\ 0 \\ \phi_1(s) \\ \phi_2(s) \\ \phi_3(s) \end{bmatrix}$$

(36)

However, as the number of actions increases, the number of parameters increases too, and more memory is required to store them. With the proposed scheme, the dimensions of $\boldsymbol{\theta}$ are in the order of $o$ ($x^M \times (L)$ *number of V-F levels × number of cores*). As such, the proposed DVFS-Enabled scheme is only feasible for systems with a moderate number of actions.

### 4.2.2 Proposed (IR)

In this section, we attempt to further reduce the complexity of our approximate $Q$-learning algorithm. At the cost of reducing the action space to only choosing a processing core (but not being able to set its working V-F level); i.e.,

$$a_k \in \mathcal{M} = \{m_1, m_2, m_3, \ldots, m_m, \ldots, m_M\} \tag{37}$$

In our proposed IR scheme, only 4 simple features are extracted for each state-action pair. In particular, each state-action pair of the system is featurized in the form of the quadruple below:

$$\{c_a, d_a(Center), d_a(Hotspot), PairingRatio_a\} \tag{38}$$

where, we have:

- $c_a$: the sensor reading associated with the chosen core $a$.
- $d_a(Center)$: the Euclidean distance of core $a$ from the chip center. The rationale for including this feature is that cooling the cores located at the center of the chip is harder if they become hot.
- $d_a(Hotspot)$: the Euclidean distance of core $a$ from the hottest on-chip component. In fact, core $a$ is thermally affected by the hot adjacent cores. Moreover, assigning tasks to neighboring cores can increase the temperature of core $a$.
- $PairingRatio_a$: The ratio of data transfer paths passing through the hottest on-chip component to the number of tasks likely to be paired with the task running on the core $a$.

In our proposed IR scheme, some feature functions are related to the actions while others are based on the states. Similarly to our DVFS-Enabled scheme, we have features concerning the temperature of the cores, but in the IR scheme, we only consider as a feature the temperature of the selected core with action $a$. Given that all elements of the above feature quadruple has continuous real-valued nature, we define an RBF for each feature element. In particular, we use $\boldsymbol{\phi}^{c_a}(s,a)$, $\boldsymbol{\phi}^{d_a(Center)}(s,a)$, $\boldsymbol{\phi}^{d_a(Hotspot)}(s,a)$ and $\boldsymbol{\phi}^{PairingRatio_a}(s,a)$ to denote the thermal features, distance from the chip center, distance from the hottest component and the ratio of pairings, respectively. Also, let $x_i, i = \{1,2,3,4\}$ be the number of RBF centers used for our feature quadruple. The dimensions of the parameter vector $\boldsymbol{\theta}$ in our proposed IR scheme would be in the order of $O$ ($\prod_i x_i$), which is much less than the proposed DVFS-Enabled scheme. Hence, as the number of RBFs increases, the number of learning parameters would still be manageable.

Table 4 summarizes the notations used in our proposed learning algorithms. Also, Algorithm 1 is a generic pseudo-code outlining the overall learning procedure for both our proposed variations.

TABLE 4
NOTATIONS USED IN THE PROPOSED LEARNING ALGORITHMS

| Notation | Description |
|---|---|
| $\rho^\pi$ | The long-term average "one-stage" reward obtained by following policy $\pi$ starting from any state $s \in \mathcal{S}$ |
| $V^\pi(s)$ | The value of state $s$ obtained by following policy $\pi$ |
| $V^*(s)$ | The maximum value of state $s$ |
| $\rho^*$ | The optimal long-term average "one-stage" reward |
| $Q(s,a)$ | Value of action $a$ at state $s$ |
| $k$ | Stage index |
| $r(s,a,s')$ | Instant reward obtained by doing action $a$ in sate $s$ and transit to state $s'$ |
| $(s^*,a^*)$ | An arbitrarily chosen reference state-action pair |
| $t(s,a,s')$ | The random transition time from state $s$ to $s'$ by performing action $a$ |
| $\alpha^k$ | Learning rate |
| $\theta$ | Vector of feature weights |
| $\theta_i^a$ | Parameter vectors of action $a$ |
| $\phi_i(s)$ | The $i_{th}$ feature function |
| $g = x^M$ | Number of RBFs |
| $\sigma$ | Width of RBF features |
| $\omega$ | M-element vector, consisting of feature centers |
| $x$ | Number of feature centres |
| $c_a(t)$ | The obtained temperature from the sensor placed next to the $a_{th}$ core at time $t$ |
| $d_a(Center)$ | The Euclidean distance of core $a$ from the chip center |
| $d_a(Hotspot)$ | The Euclidean distance of core $a$ from the hottest on-chip component |
| $PairingRatio_a$ | The ratio of data transfer paths passing through the hottest CPU component to the number of tasks likely to be paired with the task running on core $a$ |
| $\phi_i^c$ | Thermal features |
| $\phi_i^{d(Center)}$ | Distant from the center of the chip |
| $\phi_i^{d(Hotspot)}$ | Distant from the hottest on-chip point |
| $\phi_i^{PairingRatio}$ | Ratio of the number of communications |

## 5 SIMULATION EXPERIMENTS AND PERFORMANCE EVALUATION

In this section, we first describe our simulation flow, settings, and tools. Then, we compare our simulation results with related previous approaches.

### 5.1 Simulation platform and settings

Here, we introduce our simulation tools and explain the role that each tool plays in the simulation workflow. Also, we show the overall workflow and specify the values for key parameters.

### 5.1.1 Simulation tools and workflow

The simulation workflow is comprised of two phases (see Fig. …): online and offline. The offline phase is performed only once at the beginning of the simulation to collect data for the online phase. The online phase, however, is repeatedly executed in successive trials.

1) More specifically, in the offline phase, we carry out the following three step**s:**

2) A multi-core processor is simulated using Sniper [37]. As our processor is made up of similar processing cores, we only create one core and use it to simulate the entire processor. We also use the Splash-2 benchmark suite to simulate various programs with different input sizes (i.e., each Splash-2 program is considered as a single task). The execution of each task in each V-F level is simulated using Sniper [۳۷].

3) We used McPAT [۳۹] simulator to calculate the static and dynamic powers of running each task at each V-F level. In particular, McPAT is fed with simulation results produced by Sniper, including the count of access to ram, main memory, etc. Next, the routers' power consumption is calculated using DSENT [40] which is fed by NoC router specifications and data injection rate.

4) In the third step of the offline phase, the floor plan of a core is simulated using HotFloorPlan [17], which takes as in input the area of the various units of a processing core (as produced by Sniper).

**Algorithm 1 The Proposed RL-based Task Scheduling Algorithm**

**Initialization**:
$k = 0;\ \theta_0 \leftarrow 0;\ \forall(s,a): Q_0(s,a) \leftarrow 0;$ Set $\hat{Q}(s^*, a^*);\ \pi: \varepsilon - greegy;$
**begin**
  **case** (*event*) **do**
    TASK_ARRIVAL:
      // an idle core exists
1:     **if** $\exists m \in \mathcal{M}\ s.t.\ b_m(k) = 0$ **then**
2:       Call Assign_Task();
3:     **end if**
    TASK_DEPARTURE:
      // the ready queue is not empty
4:     **if** $q(k)\ != 0$ **then**
5:       Call Assign_Task();
6:     **end if**
**end**
**Function** Assign_Task()
  **begin**
7:  Read temperature values $c(k)$ from temperature sensors in state $s_{k+1}$;
8:  Use (…) to calculate $r(s_k, a_k, s_{k+1})$;
9:  Obtain $t(s_k, a_k, s_{k+1})$;
10: Use (…) to calculate $Q_{k+1}(s_k, a_k)$;
11: Use (…) to calculate $\hat{Q}(s_k, a_k)$;
12: Use (…) to update $\theta$;
13: $a_k \leftarrow$ Choose an action according to policy $\pi$;
14: Apply $a_k$;
15: $s_k \leftarrow$ Current system sate;
16: Calculate $\phi(s_k, a_k)$ (…);
**end**

5) As mentioned before, the offline phase is executed only once at the beginning of the simulation. Below, we discuss the 6 steps associated with the online phase:

6) Various tasks with different input sizes arrive randomly at the system over time (according to a Poisson process). The incoming tasks are queued in the waiting line.

7) The scheduler assigns a task from the waiting line to an idle core and sets its working V-F level.

8) The newly assigned task, pairs randomly with a running task, which is not currently paired. Following a pairing, the involved NoC routers in each pairing are injected with data of an injection rate randomly chosen from [0,1].

9) The execution time of the newly assigned task is extracted from the table produced by Sniper. Also, according to the uncertainties related to transmission time and the number of involved NoC routers, each pairing's communication time is not deterministic. To capture these uncertainties, the communication time is considered as an exponentially distributed variable with rate of $\xi$.

10) The power consumption of each unit and the floor plan of a core is given to HotSpot, which generates the

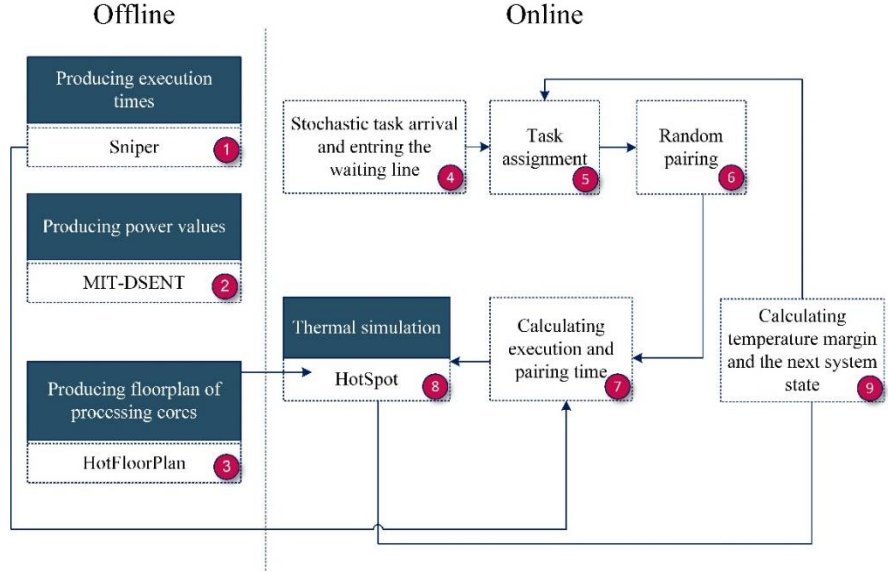

Fig. 4. Simulation workflow

11) thermal profile of the chip considering the thermal interactions of neighboring units.
12) The temperature margin is calculated and reported as an instantaneous reward to the scheduler.

### 5.1.2 Parameters and settings

The multicore processor is based on mesh architecture in which each core is an Intel Xeon Gainestown (Nehalem-EP) [51] and both NoC routers and cores are in 45 nm technology (core and router settings and V-F levels are described in Table 5 [52]). In the experiments, we vary the number of cores from 3×3 up to 7×7. The injection rate of the involved NoC routers in each pairing is randomly selected from [0,1], and the well-known xy routing algorithm [44] is used for packet routings.

TABLE 5
THE SETTINGS FOR CORES, NOC ROUTERS AND V-F LEVELS

| V-F levels [52] | | Router settings [18] | | Core settings [18] | |
|---|---|---|---|---|---|
| Voltage (V) | Frequency (GHz) | Ports | 5 | L1-I | 16 KB |
| 0.9 | 2.7 | Frequency | 2 GHz | L1-D | 16 KB |
| 1.0 | 3.0 | Virtual channels | 8 | L2 | 256 KB |
| 1.1 | 3.3 | | | ITLB | 16 entries |
| 1.2 | 3.6 | Flit size | 144 b | DTLB | 16 entries |
| | | Buffer size | 24 flits | | |

As mentioned before, each Splash-2 program is considered as a single task. The suite consists of 14 programs and each program can be executed with different input sizes. By varying these sizes, we generate a total of 29 types of tasks. Each task type enters the system with the mean Poisson rate of 0.29 tasks per second (i.e., the overall rate λ is 8.41 tasks per second for all tasks). The communication time for each pair is exponentially distributed with mean ζ. To capture the temperature of processing cores at time $t$ ($c(t)$), we implement a thermal sensor next to each core and the thermal values are obtained by reading these sensors. As the dimension of $c(t)$ corresponds the number of processing cores, implementing sensors next to every single core is not practical for larger processors. As a remedy, we use linear interpolation [۵۳] to decrease the number of temperature values to 9 irrespective of the multicore size. In linear interpolation, we exploit the known function values to fill the void of the unknowns. More formally, if function Z is given in 4 points, then the linear interpolation of this function is as: $Z(x, y) = ax + by + cxy + d$. Then, equivalence will be made for each given point and the coefficients ($a, b, c, d$) will be calculated. Using these coefficients, the value of the function can be calculated in any new point ($p,q$). By exploiting linear interpolation, we have only 9 temperature values to build $\phi$ using feature function. Also, it is considered to have 2, 3, and 5 temperature centers for each basis function, which are normalized to [0, 1] in Table 6 Therefore, depending on the number of temperature centers, we have $2^9$, $3^9$, and $5^9$ feature functions to estimate the Q-value of action $a$ in state $s$. Initial values of Q and θ are zero, and σ is defined as in table 6 depending on the number of temperature centers.

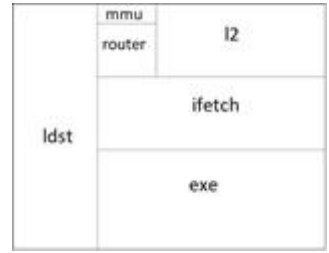

Fig. 5. The floorplan of a single processing core

TABLE 6
NUMBER OF TEMPERATURE CENTERS AND FEATURE WIDTH

| | feature width for number of centers | Temperature centers |
|---|---|---|
| 2 centers | 0.09 | 0.33 and 0.66 |
| 3 centers | 0.07 | 0.25, 0.5 and 0.75 |
| 5 centers | 0.05 | 0, 0.25, 0.5, 0.75, 1 |

Finally, in equation (27) where $a$ (the learning rate) is being calculated, A and B are considered to be 50 and 1000 respectively. The term "Iteration" refers to the repetition number of the learning algorithm. Table 7 summarizes the simulation parameters.

## 5.2 Comparison with related work

In this section, we compare our proposed algorithms with previous work:

- **TBO:** The TBO scheme in [26] has thermally managed CMPs in which a finite state machine is defined for executing multi-threaded applications. The states are named: start, wait, read, calculate, and assign. In the start state, all variables are initialized, then the algorithm switches to the wait state where the system waits for a new time quota to run a new application. In the read state, a new time quota has been started, and temperature values are collected from the on-chip thermal sensors. Then in the calculate state, a weight matrix is created such that each element of the Matrix is the inverse of the Euclidean distance of each core from the center of the chip. Then, a utilization matrix is made up of utilization values of processing cores, and it is multiplied with weight matrix to produce a cost matrix. If two cores have the same utilization simultaneously, the core closer to the center of the chip will have a higher temperature than the other. This is why TBO exploits the weight matrix in assignment decisions besides utilization and temperature values. Finally, the algorithm switches to the assigned state and allocates tasks to cores according to minimal cost principle: If two cores have the same weights, the core that has less temperature is selected to run a task.

TABLE 7
Parameters and notations

| Notation | Description | Value | |
|---|---|---|---|
| $M$ | Count of processing cores and NoC routers | 3×3, 4×4, 5×5, 6×6 and 7×7 | |
| $\lambda$ | Accumulative task arrival rate | 8.41 Tasks per Second | |
| $\zeta$ | Exponential rate of pairing time | The average execution time of paired tasks | |
| $l_L$ | Working voltage and frequency level of the processing core | 0.9/2.7, 1.0/3.0, 1.1/3.3, 1.2/3.6 (V(v)/F(GHz)) | |
| $T_{th}$ | Valid temperature threshold | 358 Kelvin | |
| $\alpha$ | Learning rate | A/(B+k) | |
| $A$ | Constant value in learning rate equation | 50 | |
| $B$ | Constant value in learning rate equation | 1000 | |
| $\sigma$ | Feature width | 2 centers | 0.09 |
| | | 3 centers | 0.07 |
| | | 5 centers | 0.05 |
| $\omega$ | Feature centers | 2 centers | 0.33, 0.66 |
| | | 3 centers | 0.25, 0.5, 0.75 |
| | | 5 centers | 0, 0.25, 0.5, 0.75, 1 |
| $z$ | Count of RBFs | 2 centers | 512 |
| | | 3 centers | 19683 |
| | | 5 centers | 1953125 |

- **LDT & LCT:** A more closely related work to our proposed scheme is [18] in which the assignment problem is formulated as an MDP [32]. Each system state consists of the temperature values obtained from the on-chip thermal sensors, and an action refers to assigning a task to a core. The instantaneous reward is the temperature margin. The work in [18] exploits reinforcement learning to solve the MDP. However, since there is no rigorous formulation of the problem, there is an ambiguity with respect to the nature of time: on the one hand, the authors have mentioned that a task is allocated to an idle core upon each task arrival. This is indicative of a continuous-time MDP. On the other hand, the Bellman equations in [18] are all based on a discrete-time MDP formulation, which contradicts the event-based operation discussed by the authors. As such, we simulate [18] in both continuous- and discrete-time versions. We use LCT to refer to the continuous-time simulation of [18], and LDT for the discrete-time version.
- **RAND:** The fourth baseline is purely random task assignment, which has very low cost of implementation, but still can serve as a standard for performance comparisons.

Our proposed DVFS-Enabled scheme works in a continues-time fashion as it assigns a task to a core and sets its V-F level as soon as a system event occurs. These events include: 1) arrival of a new task and 2) departure of a task that has been completely executed. The proposed IR scheme is similar to DVFS-enabled with the only difference that it does not apply DVFS to the cores.

## 5.3 Performance evaluation Criteria

We evaluate the proposed schemes in terms of the following criteria:

- <u>Average peak temperature</u>: The primary criterion is to reduce the average peak temperature of the processor in the long run.
- <u>Average service time</u>: the secondary criterion is service time which refers to the interval between the arrival of a task and its completion. It includes both execution time and the time that tasks wait for its turn inside the waiting line. Therefore, lower task service time means less waiting time or quicker execution, which leads to higher system performance.
- <u>The convergence of learning parameters</u>: the learning algorithm is repeated for each trial and updates its parameters and the decision making policy. The algorithmic convergence is important for system stability.
- <u>Dynamic power consumption</u>: As a peripheral goal, reducing the dynamic power (which depends on the working voltage and frequency level of the cores) is much desirable due to power budget limitations.

## 5.4 Experiments and simulation results analysis

In this section, the proposed approaches (DVFS-enabled and IR) are compared with TBO, LCT, LDT, and RAND.

### 5.4.1 Test 1: Convergence of the proposed learning algorithms

In our proposed learning algorithms, $Q$ and $\theta$ are updated at the end of each trial till they converge to their expected value in the long run. Figure 6 shows *Q*-value convergence of DVFS-Enabled for four different actions in a specific state (0.4, 0.6, 0.6, 0.6, 0.6, 0.6, 0.6, 0.6, 0.6) where only the interpolated temperature values are indicated. Figures 7 and 8 also show *Q*-value and *θ*-value convergence for our IR scheme, respectively. In Figure 7, *Q1*, *Q2* and *Q3* are *Q*-values of some randomly selected actions in states (0.1, 0.5, 0.5, 0.0), (0.5, 0.5, 0.5, 0.0) and (0.8, 0.5, 0.5, 1.0) and. Also Figure 9 indicates *θ*-value convergence for Proposed (DVFS-Enabled) and in Figures 8 and 9, Theta1, Theta2 and Theta3 are the coefficients of the first, second and third radial basis functions.

The learning rate is near to 1 at the beginning steps of learning, giving the new observations a greater effect on updating *θ* and *Q*-values. Actions are also randomly selected at the initial trials, and as a result, *θ* and *Q*-values experience high fluctuations in early stages. In intermediate trials, variations taper off, with no significant changes in final trials. This is due to the gradual reduction in the learning rate and its approach to zero. In final stages, *θ* and *Q* mostly keep their current values and take much less influence from new observations.

### 5.4.2 Test2: Average peak temperature comparison

Figure 10 shows the average peak temperature for each method. As can be seen, the proposed DVFS-Enabled scheme outperforms others. Its average peak temperature is 3 °*K* lower than LDT and LCT, 1.5 °*K* lower in comparison with TBO and as much as 6 °*K* lower than RAND. As with the proposed IR scheme, its average peak temperature is about 1.4 °*K* lower than LDT and LCT, and about 4 °*K* lower compared to RAND.

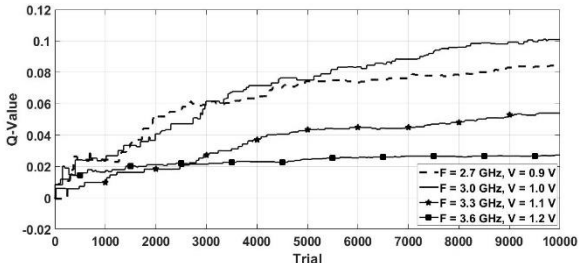

Fig. 6. Q-Value convergence of Proposed (DVFS-Enabled)

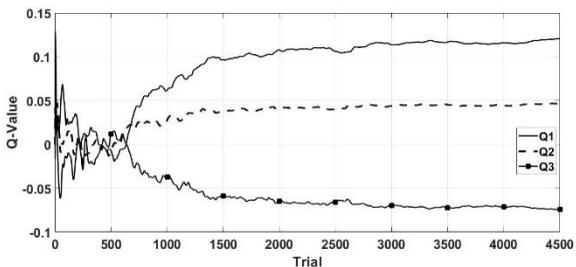

Fig. 7. Q-Value convergence of Proposed (IR)

As mentioned earlier, TBO creates a cost matrix for each core and selects a core with the minimum cost to determine the core to assign a task. This cost matrix is made from the utilization history of each core and its inverse Euclidean distance from the center of the chip. Proposed(IR) also considers the distance from the center of the chip and the hotspot, but the difference is that the proposed(IR) ignores the utilization history which may be the reason for the lower peak temperature obtained from TBO compared to the proposed (IR).

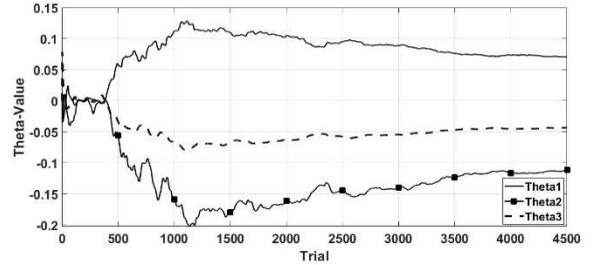

Fig. 8. θ convergence of Proposed (IR)

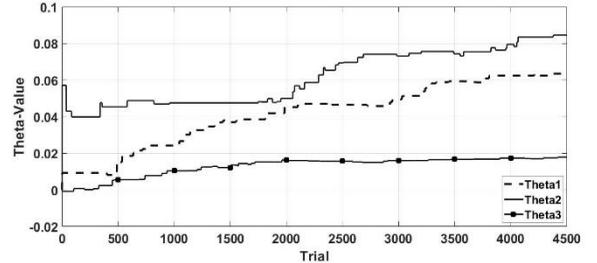

Fig. 9. θ convergence of Proposed (DVFS-Enabled)

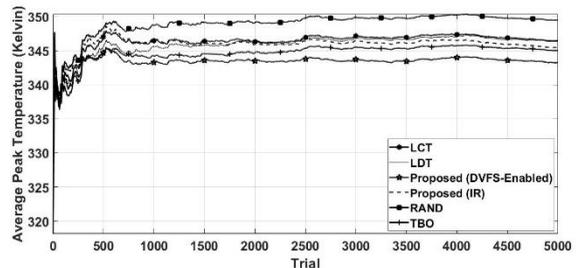

Fig. 10. The average peak temperature

### 5.4.3 Test3: Average service time

In Figure 11, both of the proposed approaches outperform LDT in terms of the average service time. In fact, LDT operates in regular time intervals without any regard to the arrival or departure events. As such, it always waits for a new time quota to assign a new task. Hence, the waiting line may remain crowded, despite the availability of several idle processing cores. The proposed schemes, on the other hand, work in a continuous-time manner, and they assign a new task as soon as an event (a task arrival or departure) occurs. Tasks are immediately assigned to cores (if the line is not empty) until there is no idle core and the overall task service time will be reduced in comparison with LDT. As a side note, since the DVFS-Enabled scheme may apply lower V-F levels to reduce the peak temperature, its task execution time is slightly higher compared to LCT, RAND, and IR. It is also noteworthy that the curves related to LCT, RAND and IR coincide in Figure 11. The reason is that all these schemes do not apply DVFS, but use a fixed and same V-F level for all the cores during the simulation.

### 5.4.4 Test4: The impact of mesh size on the average peak temperature

Figure 12, depicts the impact of mesh size (i.e., the num-

ber of processing cores) on the average peak temperature. As the mesh size increases, the peak temperature also increases according to Figure 12. Also, at the larger mesh sizes, DVFS-Enabled is outperformed by both our proposed IR and the TBO scheme. In fact, both TBO and IR avoid assigning tasks to cores near the center of the chip, and instead prefer the cores located far from the center like the edges or corners. As fewer cores surround these cores, they are less affected by the neighboring cores and are also less likely to become hot.

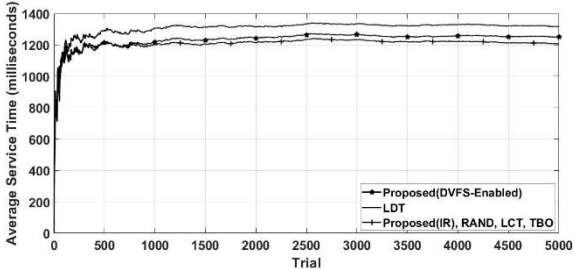

Fig. 11. The average task service time

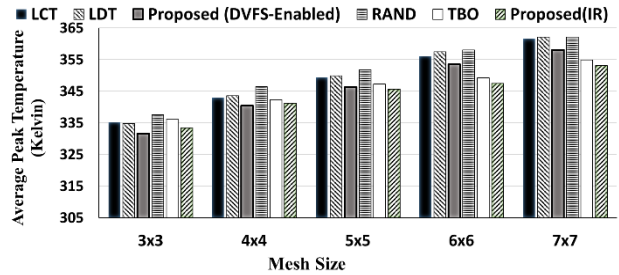

Fig. 12. the average peak temperature vs. mesh size

### 5.4.5 Test 5: Impact of the mean task arrival rate on average peak temperature

According to Figure 13, in general, the peak temperature rises with increase in the mean task arrival rate. Both the proposed schemes outperform the others for every arrival rate. In fact, IR tries to assign new tasks to the cores located far from the chip center and the hottest core. Also, DVFS-Enabled exploits several V-F levels to maximally reduce the average peak temperature. However, under higher arrival regimes, a larger number of tasks run in the system and the heat arising from the neighboring cores may affect the peak temperature. The proposed IR scheme considers the distance from the hottest core and chip center, and it partly prevents the thermal interactions among the processing cores. This is why IR outperforms DVFS-Enabled under higher arrival intensities.

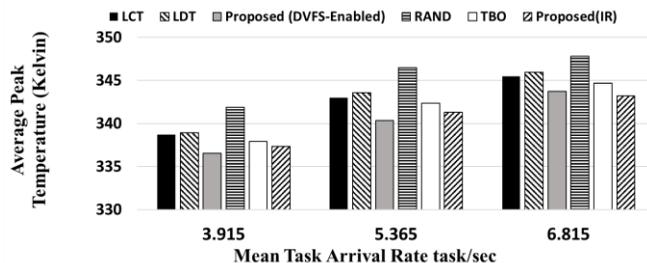

Fig. 13. The average peak temperature vs. arrival rate

### 5.4.6 Test 6: Dynamic power consumption

As mentioned before, there is an NoC router next to each processing core in a mesh topology. Each core and its local router is called a "Tile". Here, we compute the average dynamic power consumption for each tile individually in Figure 14. Using TBO and IR, the internal cores (5, 6, 9, 10) consume less dynamic power while core numbers (0, 3, 12, 15) dissipate more dynamic power. This is because of the preference of TBO and IR towards selecting cores away from the chip center (core numbering is started from 1 at the top left of the mesh to the bottom right core. The cores in each line are numbered from left to right).

The dynamic power consumption depends on the working V-F level of the processor, and since DVFS-Enabled applies DVFS to processing cores separately, this scheme is more power-efficient compared to others across all tiles. However, according to Figure 14, under DVFS-Enabled, the average dynamic power consumption is higher in internal cores as tasks may be assigned unevenly; for example, DVFS-Enabled has assigned 641 tasks (on average) to core 6 while only 108 tasks (on average) has been assigned to core 16.

Figure 15 demonstrates the total dynamic power consumption of all tiles under all approaches. As we mentioned in previous subsections, all methods except proposed DVFS-Enabled apply fixed V-F levels. Here, we set the working V-F level for these methods to 1.1 Volt and 3.3 GHz. On the other hand, there is proposed DVFS-Enabled that applies various V-F levels from Table 5. Since dynamic power depends on both voltage and frequency, the results in Figure 15 are the same for all methods except Proposed (DVFS-Enabled). Proposed (DVFS-Enabled) dissipates less dynamic power in comparison with the others, because sometimes it applies less voltage and frequency levels to reduce the average peak temperature.

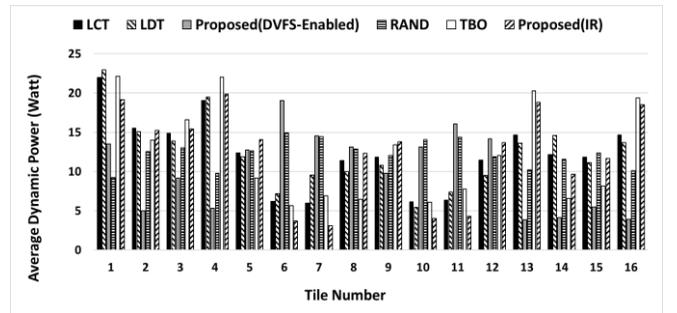

Fig. 14. Time Dynamic power consumption per each tail

### 5.6.7 Test 7: The impact of thermal center count on the average peak temperature

Since part of the state space is continuous and its other elements have a broad range of change, it is not possible to store state-action values in a table. For this reason, function approximation is used instead, which requires a feature calculator for the continuous part of the state space. As discussed in Section 4.2, we have relied on standard RBF-based approximation technique [36] with several thermal centers (ω). Also, as mentioned before, temperature values collected from on-chip sensors are reduced to only 9 values using bilinear interpolation. Therefore, $g = x^M$ features are used to estimate the value of an action in a specific state. On the other hand, if dimensions of feature vector ($\phi$) are $g \times 1$, the dimensions of $\theta$ will be the count of actions in a state $\times\ g$. As a result, an increase in the number of temperature centers enlarges the dimension of $\theta$ to an unmanageable size (Table 8).

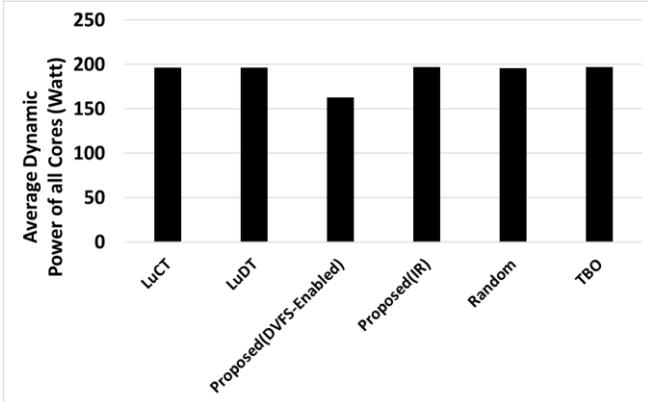

Fig. 15. Total dynamic power of tiles

In DVFS-Enabled, the count of actions is equal to 4 × number of V-F levels (L) × number of processing cores (M). In IR, the action count is reduced to M. Therefore, $\theta$ is larger in DVFS-Enabled and with an increase in temperature centers, more storage space is needed. As shown in Table 8, if there are 5 temperature centers for DVFS-Enabled, $\theta$ becomes very large, and it is infeasible to do the assignment simulations using an ordinary computer. However, in IR this issue is handled with the idea of combined state-action features such that the feature function includes features for actions besides thermal features, which results in much smaller $\theta$. Also, in IR, a higher number of temperature centers can be used to obtain higher accuracy.

TABLE 8
The count of parameters per temperature centers

|  | 2 Centers | 3 Centers | 5 Centers |
|---|---|---|---|
| **Proposed (IR)** | 348.32 K<br>16 | 347.07 K<br>81 | 345.74 K<br>625 |
| **Proposed (DVFS-Enabled)** | 346.97 K<br>51200 | 346.35 K<br>1968300 | Unmeasurable<br>195312500 |
| **LCT** | 350.12 K<br>12800 | 349.33 K<br>492075 | Unmeasurable<br>48828125 |
| **LDT** | 350.83 K<br>12800 | 349.95 K<br>492075 | Unmeasurable<br>195312500 |

### 5.4.8 Test 8: The impact of task arrival rate on the average peak temperature

In Table 9, the peak temperature changes are recorded for different arrival rates. As a general trend, higher peak temperatures are obtained as the arrival rate increases. To justify this, we should look deeper to see what happens when the arrival rate increases. As the name implies, arrival rate refers to the average number of tasks arrive to the system at each second. Now, imagine a large number of tasks enter the system at a short time. What happens is that the system becomes crowded and the processing cores execute tasks continuingly with no time to cool down. So each processing cores become hot individually while pairings and thermal interactions are also existed on the other side.

TABLE 9
The arrival rate impact on the average peak temperature

|  | 3.915 | 5.365 | 6.815 |
|---|---|---|---|
| **Proposed (IR)** | 337.32 K | 340.97 K | 343.21 K |
| **Proposed (DVFS-Enabled)** | 336.56 K | 340.04 K | 343.75 K |
| **LCT** | 338.72 K | 342.65 K | 345.48 K |
| **LDT** | 338.9 K | 343.56 K | 345.94 K |
| **RAND** | 341.89 K | 346.46 K | 347.77 K |
| **TBO** | 337.91 K | 342.35 K | 344.68 K |

### 5.4.9 Test 9: The Impact of mean arrival rate on service time

Table 10 shows the task service times for different arrival rates. Under higher arrival rates, the waiting time in the queue increases, which results in longer service times for all assignment schemes. LDT has the highest service time among others due to its discrete-time functionality.

TABLE 10
Service time per arrival rate

|  | 3.915 | 5.365 | 6.815 |
|---|---|---|---|
| **Proposed (IR)** | 1175 ms | 1207 ms | 1288 ms |
| **Proposed (DVFS-Enabled)** | 1218 ms | 1251 ms | 1330 ms |
| **LCT** | 1175 ms | 1207 ms | 1288 ms |
| **LDT** | 1284 ms | 1317 ms | 1395 ms |
| **RAND** | 1175 ms | 1207 ms | 1288 ms |
| **TBO** | 1175 ms | 1207 ms | 1288 ms |

Our proposed IR and DVFS-Enabled schemes have outperformed others under high arrival regimes. Also, as evidenced by the results, IR has even lower service time compared to DVFS-Enabled under higher arrival rates given that DVFS-Enabled sometimes exploits lower V-F levels for running the tasks.

### 5.4.10. Test 10: The thermal profile of the processor under the learning process

For investigating the learning process and its impact on assigning tasks, snapshots of the thermal profile of the processor are provided from before and after learning. Figure 16 shows the task assignment before the learning begins where tasks are paired as follows:
- The task running on core 4 is paired with task running on core 16
  - The involved NoC routers are 4, 8, 12, 16.
- The task running on core 6 is paired with task running on core 9
  - The involved NoC routers are 6, 10, 9.
- The task running on core 5 is paired with task running on core 8

- The involved NoC routers are 5, 6, 7, 8.

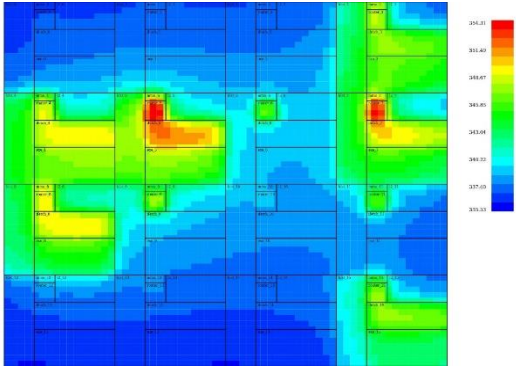

Fig. 16. Thermal profile of the processor before Learning

Figure 17 shows the thermal profile resultant from DVFS-Enabled where:
- The task running on core 13 is paired with task running on core 3
  - The involved NoC routers are 13, 14, 15, 11, 7, 3.
- The task running on core 11 is paired with task running on core 11
  - The involved NoC routers are 11 and 12.
- The task running on core 6 is paired with task running on core 8
  - The involved NoC routers are 6, 7, 8.

As shown in Figure 17, although core 7 is not assigned any tasks, it is bound on all sides by other cores, which results in its temperature raise due to thermal interactions. Thermal isolators can prevent this, but this idea is put aside due to the cost and the area occupation that it causes.

According to Figure 16, the peak temperature of the processor is 354.31K while by applying DVFS-Enabled the peak temperature is reduced to 347.4K (Figure 17). Some methods like TBO aim to reduce the peak temperature by thermal balancing, which means uniformly distributing heat to chip-wide processing cores. This is while our DVFS-Enabled scheme does not guarantee thermal balancing but has the least peak temperature.

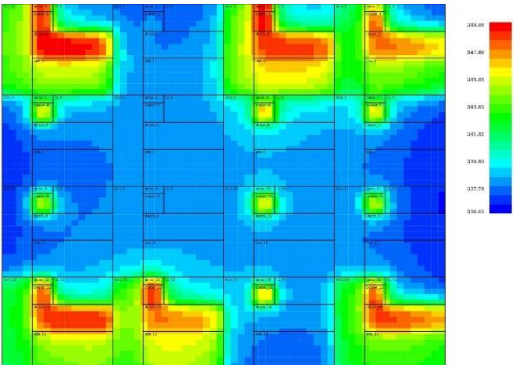

Fig. 17. Thermal profile after learning with DVFS-enabled

Figure 18 depicts the thermal profile after learning the task assignment policy by our proposed IR scheme where pairings are as follows:
- The task running on core 16 is paired with tasks running on core 4
  - The involved NoC routers are 16, 12, 8, 4.
- The task running on core 1 is paired with task running on core 13.
  - The involved NoC routers are 1, 5, 9, 13.
- The task running on core 14 is paired with task running on core 3
  - The involved NoC routers are 3, 7, 11, 15, 14.

As can be seen in Figure 18, IR tries to assign tasks to cores located away from the chip center and the hottest core. Therefore, IR can almost prevent thermal interactions.

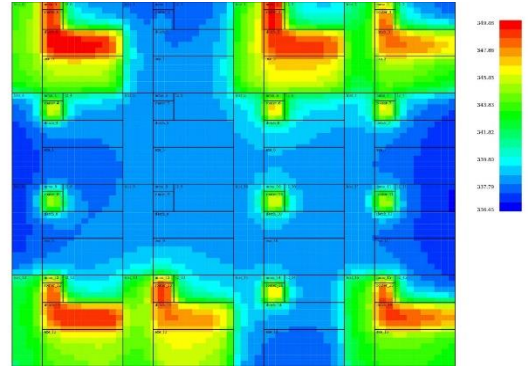

Fig. 18. Thermal profile after learning with Proposed (IR)

## 6 CONCLUSION

In this paper, we first discussed how increasing temperature affects multicore processors and how important thermal management is. Then, we reviewed several surveys and some recent papers which thermally manage CMPs. Based on our review of previous work, we categorize methods on two groups including batch and online that each related method falls into a certain category based on the presence of arrival tasks at the scheduling time. Also we mentioned that in real-world scenarios, the system (as an environment for a learning agent) is not deterministic, and includes several uncertainties which play a significant role in determining the future state of the system. These uncertainties include arrival, workload characteristics, pairings and thermal interactions. Arrival uncertainty is violated in batch methods since whole tasks are entered the system prior to scheduling. However, the online group also is potential to consider all types of uncertainties. This group is also divided into two subgroup: discrete-time and continuous-time. Discrete-time methods wait for new time quota to assign a task to a processing core, even there are several idle cores. For this reason, they result in higher service-time specially when the system is crowded. Therefore, we tried to present a continuous-time method. For this propose, we used MDP platform to formulate the task assignment problem in a multicore processor and because the statistical knowledge of the environment was not available, we used a model-free learning approach to solve the MDP. In this paper, we present two continuous-time methods named: Proposed(DVFS-Enabled) and Proposed(IR). In both presented methods, a state space is defined to specify the system state at each step and the operating system scheduler is considered a learning agent that selects an action at each system state. By doing each action, the system state is changed and the scheduler earns a reward that specifies the goodness of the selected action in reaching the system goal. Here the long-term goal is defined as reducing the average peak temperature of the CMP. The reward also is used for updating the last action- and the previous state-values which are going to be used in future decision makings. Therefore, action- and state-values converge to optimal values in long-run and the optimal policy is calculated.

Our simulation results indicate that in most cases, both proposed approaches outperform the others in reducing the average peak temperature and since both methods are used to work in continuous-time manner, they result in less service time. Also, we show that proposed(DVFS-Enabled) dissipates less dynamic power in comparison with the others which is due to applying lower voltage and frequency levels to keep the average peak temperature as low as possible.

Comparing the two methods presented in this paper, Proposed(DVFS-Enabled) uses a large number of parameters for learning which increases when exploiting more temperature centers which in turn, slows down learning. On the other hand, Proposed(IR) is capable to use more centers for state features since it needs less learning parameters. Therefore, it learns faster and more precise.

Both proposed methods can be improved by adding several ideas; some are suggested below:

- Since using function approximation is problematic with off-policy and its convergence is not guaranteed, policy gradient can be used instead.
- Several constraints such as length of the waiting queue and the number of involved routers in pairings can be considered to increase the performance, reduce the average peak temperature or energy efficiency.
- Include other system events; for example, exceeding a predetermined temperature threshold. This event occurs when the temperature of a processing core goes beyond the threshold. In this case, the operating system scheduler migrates the running task to an idle core and applies DPM techniques to cool down the hot core.